\documentclass[aps,pra,preprint,groupedaddress,bibnotes,citeautoscript]{revtex4-1}

\usepackage{graphicx}
\usepackage{dcolumn}
\usepackage{bm}
\usepackage{color}

\begin{document}


\title{Tree based machine learning framework for predicting ground state energies of molecules}

\author{Burak Himmetoglu}
\affiliation{Center for Scientific Computing, University of California, Santa Barbara, CA 93106}
\affiliation{Enterprise Technology Services, University of California, Santa Barbara, CA 93106}

\date{\today}

\begin{abstract}
We present an application of the boosted regression tree algorithm for predicting ground state energies of 
molecules made up of C, H, N, O, P, and S (CHNOPS). The PubChem chemical compound database has been 
incorporated to construct a dataset of 16,242 molecules, whose electronic ground state energies
have been computed using density functional theory. This dataset is used to train the
boosted regression tree algorithm, which allows a computationally efficient 
and accurate prediction of molecular ground state energies. Predictions from boosted regression trees
are compared with neural network regression, a widely used method in the literature,
and shown to be more accurate with significantly reduced computational cost. The performance of the
regression model trained using the CHNOPS set is also tested on a set of distinct molecules that contain
additional Cl and Si atoms. 
It is shown that the learning algorithms lead to 
a rich and diverse possibility of applications in molecular discovery and materials informatics.
\end{abstract}

\maketitle

\section{\label{sec:intro} Introduction}

Advances in electronic structure theory in combination with ever increasing computing power have made quantum mechanical
simulations of molecules and solids rather commonplace. As a consequence, vast amounts of compounds have been
studied by various electronic structure methods. In the last decade, efforts to collect and catalog these simulation data,
in addition to those obtained from experimental studies have led to the ability to analyze and screen an extensive space 
of chemical compounds and materials~\cite{Jain:MGenome,Curtarolo:AFLOW}. 
Automated screening of the experimental and theoretical compound spaces has become a powerful
tool not only for discovering new systems, but also for rational design of chemicals and materials
for targeted applications~\cite{Curtarolo:Alloys,Hautier:Oxide,Pizzi:Aiida}. 
Moreover, with the availability of various databases, analyses based on powerful statistical/machine
learning methods have become feasible. Despite the availability of a vast number of systems in these databases, screening for new systems
which have not yet been reported before requires a large number of new simulations to be performed. 
Density functional theory~\cite{DFT-1} (DFT), based on the effective single-particle Kohn-Sham equations~\cite{DFT-2} 
has been a popular choice for performing accurate, yet computationally inexpensive simulations. Despite the relatively low cost of DFT simulations, 
screening the whole space of compounds and materials require electronic structure 
predictions at a much lower computational cost, ideally without performing new simulations for each system. 
Machine learning algorithms are a perfect match for this task, since these algorithms could learn from a given
database of electronic structure calculations, and predict the electronic properties of a new set of systems (not included in 
the database) without the need of performing new simulations. Such a task requires a set of features (a.k.a descriptors)
for each system that define their electronic makeup. Then, the electronic structure can be described by a general nonlinear function of these features,
which the learning algorithm determines by a sophisticated fit to a given database (i.e. training the learning
algorithm). As a result, predictions on new systems can be readily obtained through the trained model 
parameters. 

The idea of predicting electronic structure using data has already been investigated in the literature. Various
learning algorithms and choice of features have been proposed to predict electronic structures of molecules and solids. 
Earliest investigations considered a combined DFT and machine learning based modeling for predicting
potential energy surfaces, ground state energies and formation enthalpies of molecules. 
These studies used neural networks~\cite{Behler:nnet,Hu:nnet,Balabin:nnet,Sun:nnet}
and support vector regression~\cite{Balabin:svm} as learning algorithms.
In case of solids, features were constructed from calculated electronic band gaps, cohesive energies, and crystalline
volumes for a set of inorganic compounds and used for predicting a new set of band gaps using 
support vector regression~\cite{Lee:svr}.
In principle, apart from the approximation to the exchange-correlation functional (E$_{\rm xc}$), DFT calculations use atomic species and their positions as the only
inputs for ground state electronic structure predictions~\footnote{In practice, the situation is complicated further by the choice
of pseudopotentials, basis sets, kinetic energy cut-offs, boundary conditions (especially for molecular systems) and the simulation 
code~\cite{reproducibility}}.
In a similar manner, features for learning algorithms can be constructed
based only on atomic coordinates and types, which provide an improvement over features based on calculated properties.  
For this objective, Coulomb matrices have been proposed as a robust set of features for a complete
description of molecular systems~\cite{Rupp:CIJ}. With Coulomb matrices as features, various learning algorithms have been tested, 
ranging from nonlinear kernels to neural networks, which resulted in accurate predictions for atomization energies of molecules~\cite{Rupp:jctc}. 

Given the success of learning algorithms based on Coulomb matrices as features, it is therefore imperative to analyze
available datasets~\cite{Hill:MRS} with electronic structures determined by various DFT methods and implementations. 
For this purpose, we construct a database of electronic structure calculations based on the molecular structure data
publicly available through the PubChem Substance and Compound database~\cite{pubchem}. The electronic 
structure calculations are based on the highly scalable implementation of DFT which employs plane-waves and pseudopotentials. 
For the machine learning algorithm, we propose the adoption of boosted regression trees~\cite{Friedman:GBM}, as a more computationally 
efficient alternative to previously used methods in the literature. As a demonstration, we show that the boosted regression
trees outperform neural networks in predicting atomization energies, while significantly reducing the cost of model training.  
Our framework, based on Coulomb matrices as features and boosted regression trees as the machine 
learning algorithm, provide an accurate and efficient pathway to screen the chemical compound space
for analyzing data and discovering new molecules. 

The paper is organized as follows: We provide details about the computational methods used and the construction
of the electronic structure database in section~\ref{sec:data}. In section~\ref{sec:ml}, we provide a 
summary of the machine learning algorithms and the techniques used for model training. In section~\ref{sec:res},
we present the main results and provide a discussion on the performance of the methods used. 
Finally, in section~\ref{sec:conc}, we provide some concluding remarks.

\section{\label{sec:data} Methods}

\subsection{Obtaining Data and Computational Methods}
Molecular structures used in our study are generated from the PubChem Substance and Compound database~\cite{pubchem}.
The chemical structures which are probed have the substance identifier number (SID) ranging from 
1 to 75,000. Using the structural information from this database, a subset of molecules are extracted
based on the number and types of atoms, and the size of the molecules. This subset contains molecules
that satisfy the following criteria:
(i) Each molecule has to be composed of a subset of the elements from the set C, H, N, O, P and S (CHNOPS). 
(ii) Each molecule must have at least 2, at most 50 atoms. (iii) The maximum distance between two atoms
in a molecule must not exceed 25 a$_0$ ($a_0=$ 0.529 \AA, i.e. Bohr radius), for convergence of plane-wave calculations, 
where each molecule is placed in a cubic box of side length 30 a$_0$. 
(iv) There must be an even number of electrons in the molecule.
Applying these criteria to the first 75,000 entries in the PubChem database leads to a subset of 16,242 molecules,
whose structure data files (SDF) are converted into input files for electronic structure calculations.

The electronic structure calculations are performed using the plane-waves pseudopotential 
implementation of DFT in the PWSCF code of the {\it Quantum ESPRESSO} package~\cite{QE}. The exchange-correlation
energy is approximated using the generalized gradient approximation (GGA) with the Perdew-Burke-Ernzherof (PBE)
parametrization~\cite{pbe}. 
The choice of the PBE functional is purely due to its computational efficiency in the plane-waves basis set.
Since the aim of this work is to predict simulation results, PBE is sufficient for our purposes. Functionals with exact-exchange 
that perform better can also be used for calculation of ground state energies which can be fed into the same machine learning
algorithms used in this study.
All of the atoms are represented by ultrasoft pseudopotentials~\cite{uspp}. 
The electronic wavefunctions and charge density are expanded up to kinetic energy cutoffs of 30 Ry and 300 Ry, respectively. 
The calculations are performed in a cubic box of side length 30 a$_0$ with a single k-point at $\Gamma$.   
These choices lead to the calculation of the ground state energy with a numerical error of about 1 kcal/mol for the largest molecule
in the dataset. 

\begin{figure}[!ht]
\includegraphics[width=0.5\textwidth]{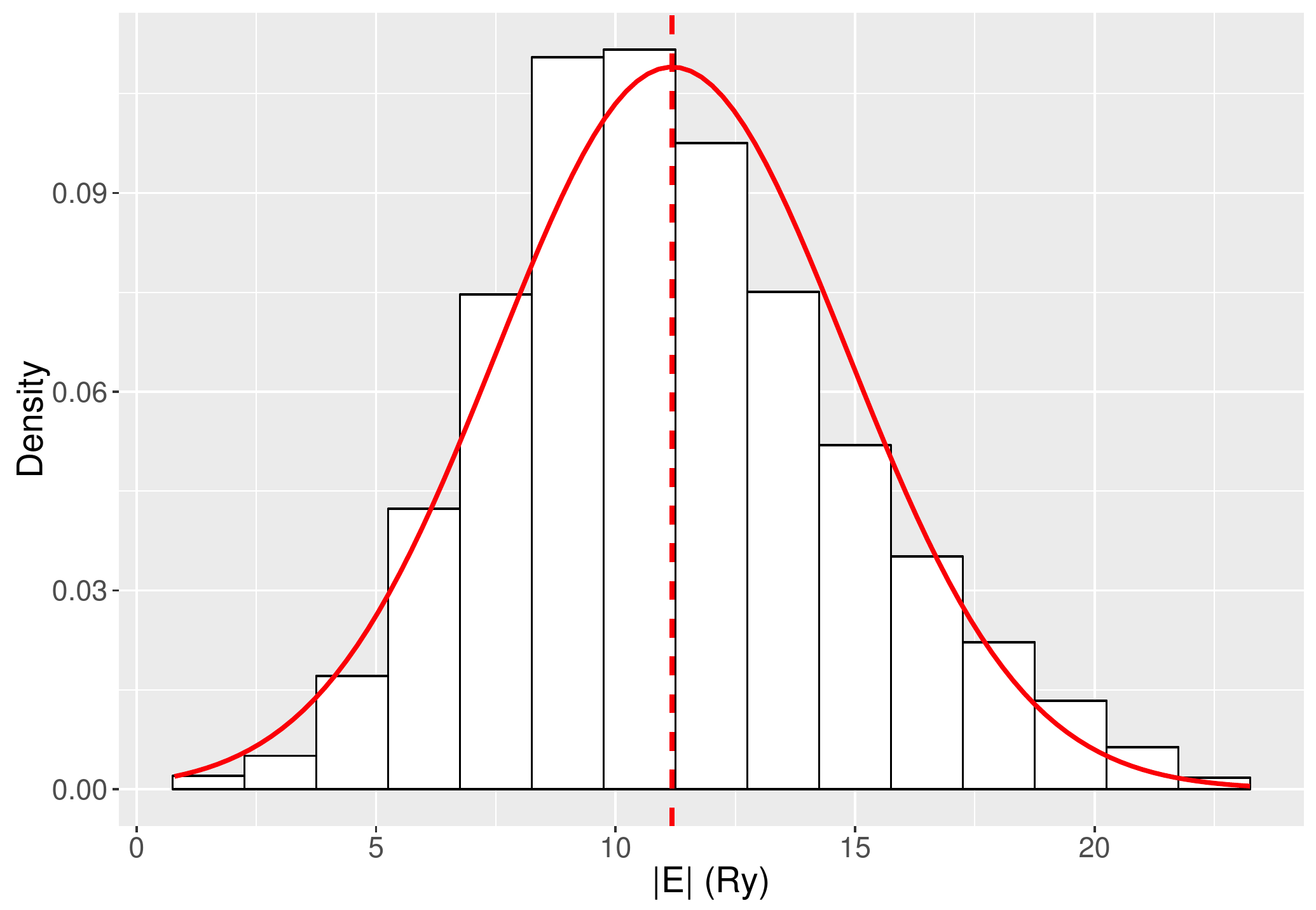}
\caption{\label{fig:AE} Histogram of pseudo-atomization energies (E$_{\rm ps}$). The mean value of 
 $\vert E_{\rm ps} \vert$ is 11.18 Ry (3506.37 kcal/mol) (indicated by the dashed vertical line) and the standard deviation
is 3.66 Ry (1147.89 kcal/mol). A Gaussian probability density function with the same mean and standard deviation is also plotted
for comparison.}
\end{figure}

For each molecule in the dataset, we compute the pseudo-atomization energy (E$_{\rm ps}$), which is the quantity
that serves as the prediction (outcome) for the learning algorithms. We compute E$_{\rm ps}$ using
\begin{equation}
E_{\rm ps} = E_{\rm gs} - \sum_{\alpha=1}^{N}\, n_{\alpha}\, E_{\rm \alpha}^{\rm PS} \label{eqn:Eps}
\end{equation}
where E$_{\rm gs}$ is the calculated DFT ground state energy, N is the number of atoms in the molecule,
the index $\alpha$ specifies the type of the atom (belonging to the CHNOPS set), n$_{\alpha}$ is the number of atoms of type $\alpha$, 
and E$_{\alpha}^{\rm PS}$ is the pseudo-energy of the isolated atom of type $\alpha$ calculated during pseudopotential
generation. All pesudopotentials are generated from the PSLibrary repository, provided in the 
{\it Quantum ESPRESSO} distribution.  
The histogram for the calculated pseudo-atomization energies from the dataset of 16,242 molecules are shown in Fig.~\ref{fig:AE}.
The total variability in the dataset, quantified by the standard deviation of E$_{\rm ps}$, is 3.66 Ry (1147.89 kcal/mol), which is larger than 
the variability reported in some of the earlier works in the literature~\cite{Rupp:jctc,Rupp:NJP}, indicating that a much wider range of molecular 
systems being included in the current study. The consequences of this difference will be discussed in section~\ref{sec:res}.

\subsection{Data Description and Visualization}
In order to build models for predictions from machine learning algorithms, construction of feature vectors (a.k.a descriptors) are needed
to represent each molecule. 
We use the intermolecular Coulomb repulsion operators (will be referred to as Coulomb matrices from now on)
introduced in Ref.~\onlinecite{Rupp:CIJ}, which are defined as
\begin{equation}
C_{IJ} = \bigg\{ \begin{array}{c} 0.5\, Z_I^{2.4} \,\,\,\,  I=J  \\ 
                 \frac{Z_I\, Z_J}{\vert {\bf R}_I - {\bf R}_J \vert} \,\,\,\, I \neq J \end{array}
\label{eqn:Cij}
\end{equation}
where Z$_I$ are atomic numbers, ${\bf R}_I$ are atomic coordinates, and indices $I,J$ run over the atoms in a given molecule. The off-diagonal terms correspond to
ionic repulsion between atoms $I$ and $J$ and the diagonal terms (both the pre-factor and power)
are obtained from a fit of the atomic numbers to the energies of isolated atoms~\cite{Rupp:CIJ}. 
The Coulomb matrices represent the full set of parameters that DFT calculations take as inputs (Z$_I$ and ${\bf R}_I$), aside from the
approximation to E$_{\rm xc}$, which are used to calculate the ground state energy. Therefore, the problem of predicting ground state energies (or atomization
energies) using C$_{IJ}$ is well defined. However, C$_{IJ}$ does not provide a unique description, since a given molecule
can be represented by more than one matrix that can be obtained by reshuffling the indices of the atoms. There are several ways to overcome this problem as described
in Ref.~\onlinecite{Rupp:jctc}. Here, aside from C$_{IJ}$ itself, we will use its eigenspectrum (which is one of the unique representations 
proposed in Ref.~\onlinecite{Rupp:jctc}) for a given molecule.   
Since we limited the number of atoms in a given molecule by 50, the Coulomb matrices we use are $50 \times 50$ matrices. Molecules with less than
50 atoms have their Coulomb matrices appended by columns and rows of 0 to complete them to have dimensions of $50 \times 50$.  

A given molecule in the dataset numbered with index $i$ is represented by a $p$-dimensional 
feature vector ${\bf x}_i$, where $p$ is the total number of unique entries in the Coulomb matrix (i.e. the upper triangular part of the symmetric $50 \times 50$ 
matrix C$_{IJ}$, unrolled into a 1275 dimensional vector) or the number of eigenvalues (i.e. 50 dimensional vector of eigenvalues). 
The whole dataset is then cast in a data matrix ${\bf X}$ 
of dimensions $N \times p$, where N is the number of data points (16,242). In this representation, molecules are listed in rows, and each column is an entry
in the p-dimensional feature vector ${\bf x}_i$ (for the $i^{\rm th}$ molecule). Namely, for a given molecule with index $i$, 
\begin{equation}
{\bf x}_i = \left[ \begin{array}{c} C_{1,1}^{(i)} \\ C_{1,2}^{(i)} \\ \vdots \\ C_{1,50}^{(i)} \\ C_{2,2}^{(i)} \\ \vdots \\ C_{2,50}^{(i)} \\ \vdots  
\\ C_{49,50}^{(i)} \\ C_{50,50}^{(i)} \end{array} \right], 
\,\,\,\, i = 1, \dots, N
\label{eqn:xC}
\end{equation}
when the full Coulomb matrix $C_{IJ}^{(i)}$ is used, while
\begin{equation}
{\bf x}_i = \left[ \begin{array}{c} \lambda_1^{(i)} \\ \lambda_2^{(i)} \\ \vdots \\ \lambda_{50}^{(i)} \end{array} \right], \,\,\,\, i=1,\dots,N 
\label{eqn:xlambda}
\end{equation}
when the eigenvalues $\lambda^{(i)}$ of $C_{IJ}^{(i)}$ are used. Finally, the data matrix ${\bf X}$ is explicitly given in terms of the 
feature vectors ${\bf x}_i$ by 
\begin{equation}
{\bf X}= \left[ \begin{array}{c} {\bf x}_1^{\rm T} \\ {\bf x}_2^{\rm T} \\ \vdots \\ {\bf x}_N^{\rm T} \end{array} \right],
\,\,\,\, i = 1, \dots, N  \label{eqn:X}
\end{equation}
The outcomes, namely the pseudo-atomization energies E$_{\rm ps}$, are also cast in a N-dimensional vector explicitly given by
\begin{equation}
{\bf y} = \left[ \begin{array}{c} E_{\rm ps}^{(1)} \\ E_{\rm ps}^{(2)} \\ \vdots \\ E_{\rm ps}^{(N)} \end{array} \right], \,\,\,\, i=1, \dots, N
\label{eqn:Y}
\end{equation}
As will be explained in the next section, the objective of the machine learning problem is to find a nonlinear function $f({\bf X})$ 
that learns a relationship between the data ${\bf X}$ and the outcomes ${\bf y}$. 

Fig.~\ref{fig:AE} depicts the overall variability in the outcome vector ${\bf y}$, namely the distribution of E$_{\rm ps}$
in the dataset. However, it is also useful to explore the features ${\bf X}$ and how they relate to E$_{\rm ps}$. This is not 
a straightforward task, since the data matrix ${\bf X}$ have a very large number of columns (1275 when data is 
represented by $C_{IJ}$, 50 when data is represented by $\lambda$). Instead of plotting E$_{\rm ps}$ as a function
of all possible features, a smaller set of features that describe the overall trends in the data can be constructed using
principal components analysis (PCA)~\cite{elsl}. PCA enables one to obtain linear combinations of features (i.e. columns of ${\bf X}$)
that explain the largest variation in the data. Before utilizing PCA, the data matrix is centered via the transformation
\begin{equation}
X_{ik} \rightarrow X_{ik} - \mu_k, \,\,\,\, \mu_k = \frac{1}{N}\, \sum_{i=1}^N\, X_{ik} \label{eqn:center}
\end{equation}
This transformation ensures that each feature has zero mean, i.e. the column sums of ${\bf X}$ are 0. 
With the centered data, the covariance matrix takes a simple form
\begin{equation}
\Sigma_{ij} = \frac{1}{N}\, \sum_{k=1}^N\, X^T_{ik}\, X_{kj} \label{eqn:cov}
\end{equation}
The principal components are constructed from the eigenvectors of the covariance matrix,
and are ordered with increasing eigenvalues. For instance, the first principal component
corresponds to the linear combination of columns of ${\bf X}$ with the largest eigenvalue, hence has the largest variance. 
Formally, the $j^{\rm th}$ principal component vector is given by
\begin{equation}
(Z_{j})_i = \sum_{k=1}^{p}\, \phi_k^{(j)}\, X_{ik}
\end{equation}
where
\begin{equation}
\sum_{k=1}^p\, \Sigma_{ik}\, \phi_k^{(j)} = \sigma^2_{(j)}\, \phi_i^{(j)}
\end{equation}
In the above equation, $\sigma^2_{(j)}$ is the eigenvalue corresponding to principal component vector ${\bf z}_j$. 
Using this recipe, we have constructed the principal components of ${\bf X}$ using the eigenspectrum $\lambda^{(i)}$ (Eqn.(\ref{eqn:xlambda})).
Among the 50 principal components, the first two account for 32 \% of the variability in the data (i.e., 
$\sigma^2_{(1)}+\sigma^2_{(2)} = 0.32\, \times {\rm Tr}[\Sigma]$).  
\begin{figure}[!ht]
\includegraphics[width=0.8\textwidth]{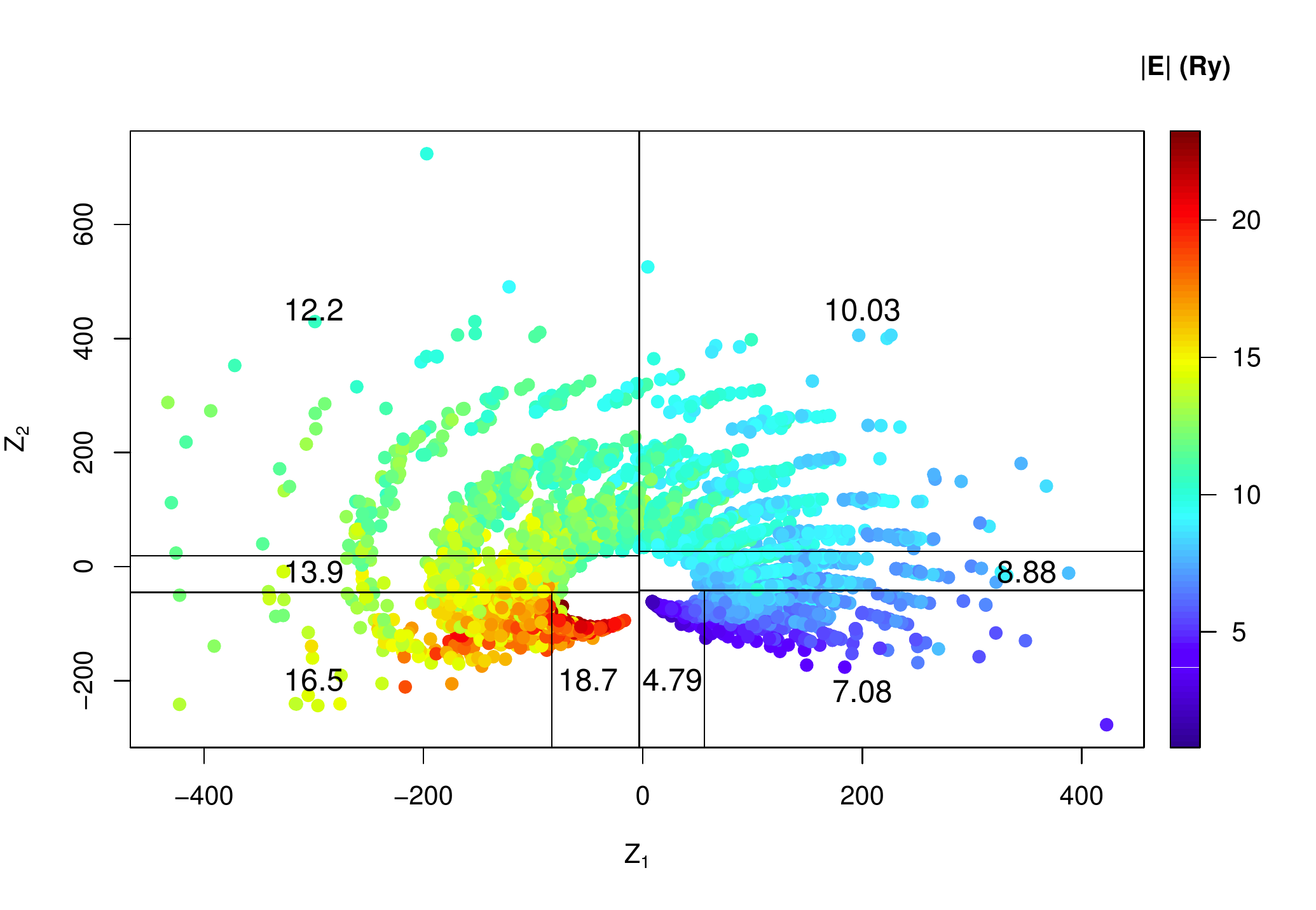}
\caption{\label{fig:PCA} E$_{\rm ps}$ as a function of the first two principal components Z$_1$ and Z$_2$. The reported values within each 
rectangular region are the mean values of E$_{\rm ps}$, which are determined by the regression tree algorithm
outlined in section~\ref{sec:ml}.} 
\end{figure}
Fig.~\ref{fig:PCA} illustrates E$_{\rm ps}$ as a function of the first two principal components ($Z_1$ and $Z_2$),
which display a peculiar nonlinear dependence on the two features. In the next section, we will summarize some of the learning techniques that
we have used to accurately model this nonlinear behavior of E$_{\rm ps}$ based on the data.

\section{\label{sec:ml} Learning Methods}

\subsection{Overview}
\begin{figure}[!ht]
\includegraphics[width=0.8\textwidth]{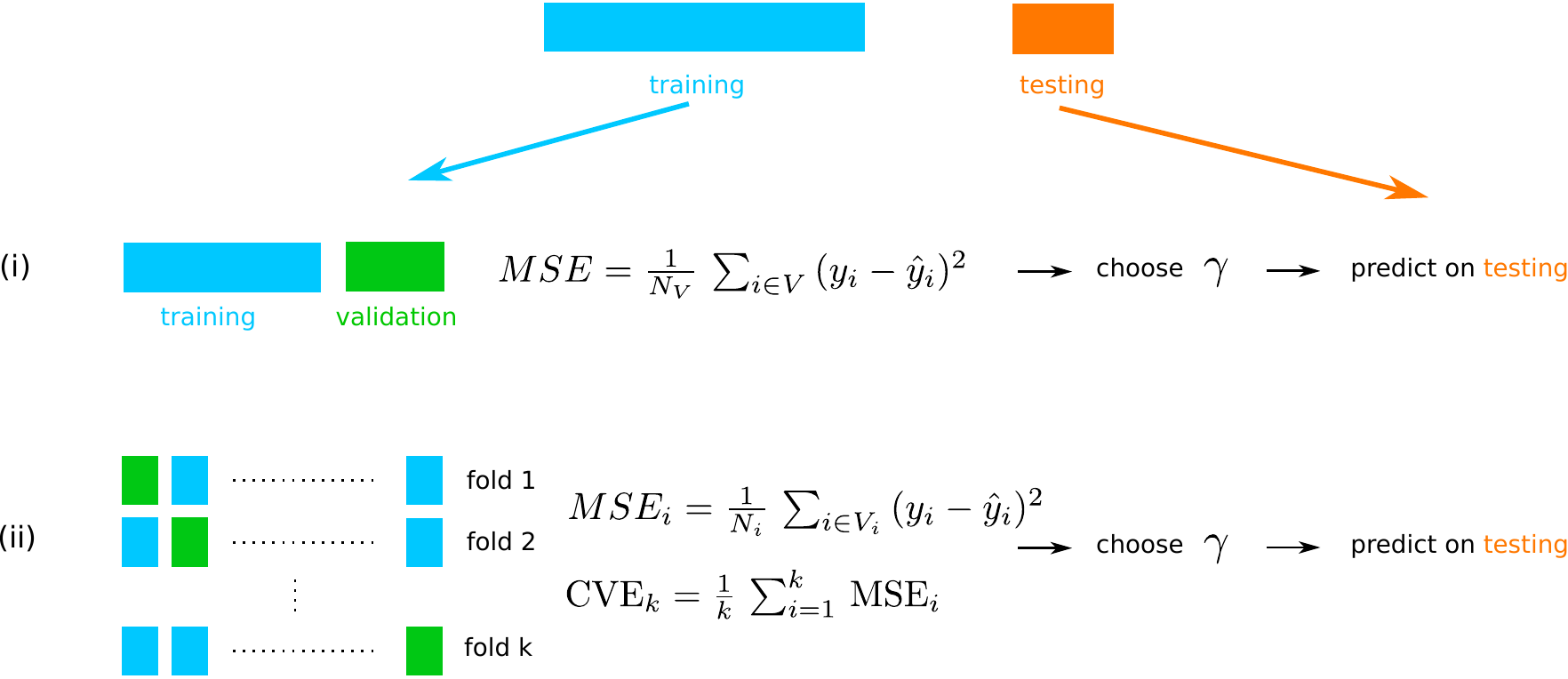}
\caption{\label{fig:CV} (i) Validation set and (ii) k-fold cross-validation approaches used for training a model. In case of LASSO,
a single regularization parameter $\gamma$ is picked from the model with lowest cross-validation error. Final accuracy is 
reported on the independent testing set.}
\end{figure}

The main purpose of training a learning algorithm in the context of regression is to fit a nonlinear
function $f({\bf X})$ using the data ${\bf X}$ to predict the outcome ${\bf y}$. This is achieved
via the minimization of a loss function which is a measure of the difference between the actual data and the fit.
For instance, in the case where $f$ is parameterized 
by a p-dimensional vector $\theta$, and the predictions of the outcomes ${\bf y}$ are given by
${\hat {\bf y}} = f_{\theta}( {\bf X} )$, $\theta$ is obtained by optimizing
\begin{equation}
\min_{\theta}\, \left\{ \sum_{i=1}^N\, \left( y_i - {\hat y}_i \right)^2 + R(\theta) \right\} \label{eqn:loss}
\end{equation}
where $y_i - {\hat y}_i$ are residuals and the loss function used for optimization is the residual sums squared (RSS).
The term $R(\theta)$ is the regularization term, and prevents overfitting~\cite{elsl}. For example, in the well-known 
case of LASSO~\cite{elsl}, a linear function $f_{\theta}({\bf X}) = {\bf X} \cdot \theta$ is used for predictions ${\bf {\hat y}}$, and
the regularization term is $R(\theta) = \gamma\, ||\theta||$. While the parameter vector $\theta$ is obtained by minimizing Eqn.(\ref{eqn:loss}),
the regularization parameter $\gamma$ is obtained by cross-validation. In this article, we consider two types of cross-validation approaches:
(i) validation set, (ii) k-fold cross-validation. In both approaches, the full dataset is randomly divided into a training and test
set. In this study, 70 \% of the data is randomly selected as the training set, and 30 \% as the test set.
The training set is used to train the learning algorithm, i.e. to obtain values of the regularization parameter(s), while the test 
set, as an independent piece of the data, is used to report the accuracy of the trained model. 
The validation set and k-fold cross-validation approaches are schematically illustrated in Fig.~\ref{fig:CV}.
In the validation set approach, the training data is further split (randomly) into two, yielding a second level training set and a validation set.
The model is trained on the training set for a range of parameters that determines the regularization term. 
Then, the model which results in the smallest mean squared error (MSE) is chosen. The MSE is the mean value
of the error in the fit given by
\begin{equation}
 {\rm MSE} = \frac{1}{N}\, \sum_{i=1}^N\, (y_i - {\hat y}_i)^2. \label{eqn:MSE}
\end{equation}
In case of LASSO, $\gamma$ is the parameter determined from cross-validation. Finally, the first level training set (the original one with 70 \% of the data) is
used to re-fit the model (with the regularization parameters fixed) and the final accuracy of the model is reported by the performance on the test set. While this method
provides a clear path to determining the regularization parameters, the results generally depend on the way the data is split.
The k-fold cross-validation approach attempts to resolve this issue by randomly diving the training set into k groups of approximately
equal size. For each division (i.e. fold), the validation set approach is repeated: the (k-1) folds are used as the second level training set and the 
left out fold is used as the validation set. After obtaining k estimates for the MSE, the results are averaged and the model which
leads to the smallest cross-validation error (CVE) is picked, as depicted in Fig.~\ref{fig:CV}. 
k-fold cross-validation usually resolves the dependence of the model parameters on the splits used, at the expense of increased
computational cost. We use 5-fold cross-validation for training learning algorithms in this study, except the case of neural networks
when the full $C_{IJ}$'s are used as features (a validation set is used in that case).

\subsection{Regression Trees}
\begin{figure}[!ht]
\includegraphics[width=0.8\textwidth]{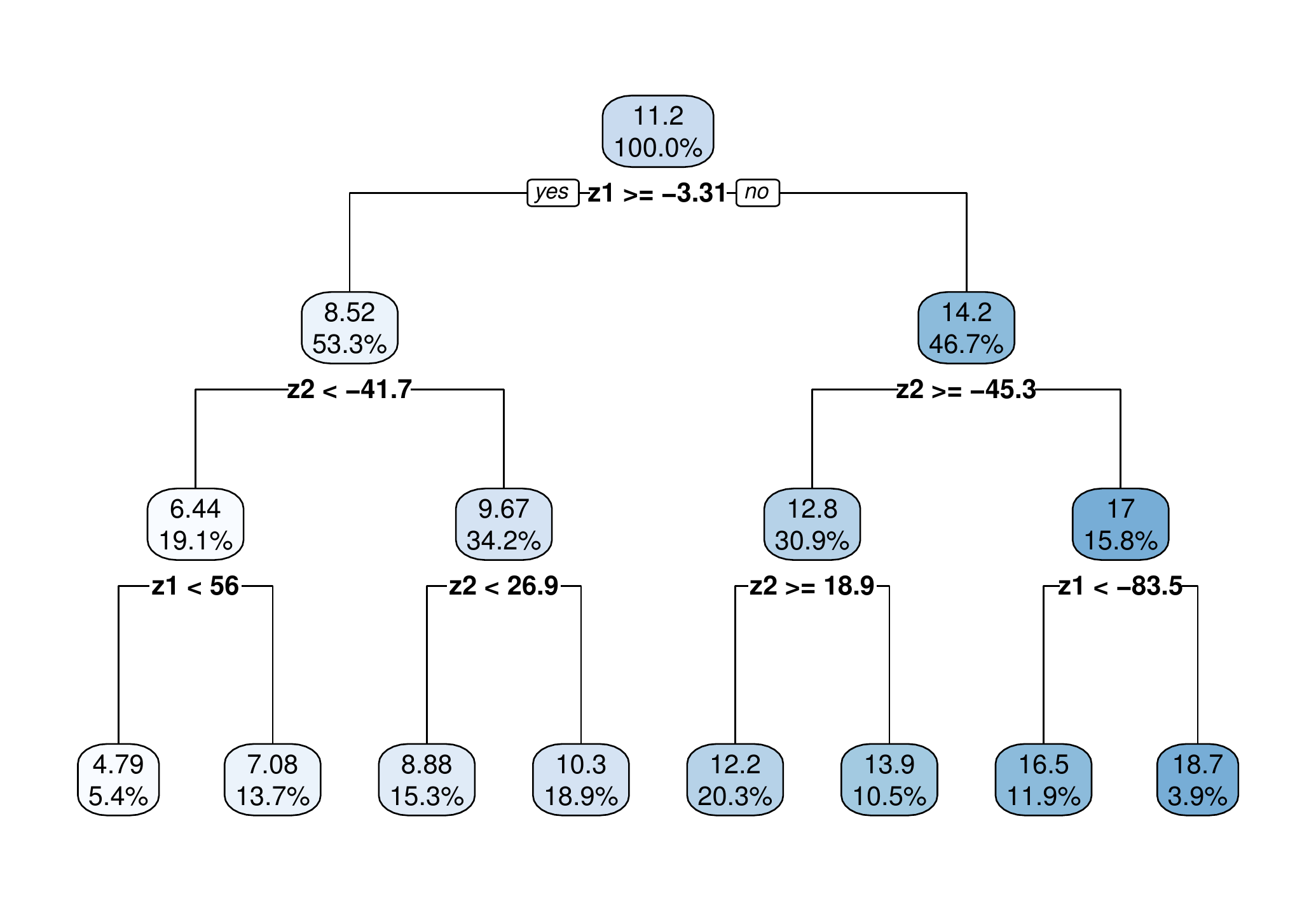}
\caption{\label{fig:rpart} Regression tree trained on the dataset with the first two principal components used
as features. The blue boxes contain the value of the average value of E$_{\rm ps}$  as
well as the percentage of data points in the region they belong to. These regions can also be seen in Fig.~\ref{fig:PCA}.}
\end{figure}

The idea behind regression trees is to divide the space of features into regions where in each region, the value of 
the function $f({\bf X})$ is the mean of the observations $y_i$ inside. For example, Fig.~\ref{fig:PCA}
illustrates the result of a regression tree fitted using two features that are the principal components $Z_1, Z_2$. In
each of the rectangular regions of the feature space spanned by $\{Z_1, Z_2\}$, the prediction for $y_i$ is the average
of the observations (values inside each rectangle in Fig.~\ref{fig:PCA}). Namely,
\begin{equation}
{\hat y}_{i \in R_J} = \frac{1}{N_J}\, \sum_{i \in R_J}\, y_i \label{eqn:treeavg}
\end{equation}
where $R_J$ is a region in the feature space and $N_J$ is the number of data points in $R_J$. 
The regions $R_J$ are determined by optimum splits in the feature space that leads to the smallest
RSS. Formally, the following function is minimized to obtain the tree
\begin{equation}
L = \sum_{t=1}^{T}\, \sum_{i \in R_t}\, \left( y_i - {\hat y}_i \right)^2 + \gamma\, T \label{eqn:losstree}
\end{equation}
where the first term is the RSS, and the second term act as the regularization. $T$ is the number of
terminal nodes (or leaves) in a tree which coincide with the number of regions the feature space
is split. The larger the number of terminal nodes $T$, the more complex the structure of $f({\bf X})$
will be, which may lead to overfitting. The regularization term adds a penalty for complex trees to prevent overfitting.
An example of a tree with 8 terminal nodes is shown in Fig.~\ref{fig:rpart}.
As can be seen, the regression tree is grown recursively. The largest reduction in RSS is obtained at the split $Z_1=-3.31 $,
then in splits $Z_2 = -4.17\, , \, Z_2 = -45.3$, and so on.

\begin{figure}[!ht]
\includegraphics[width=0.8\textwidth]{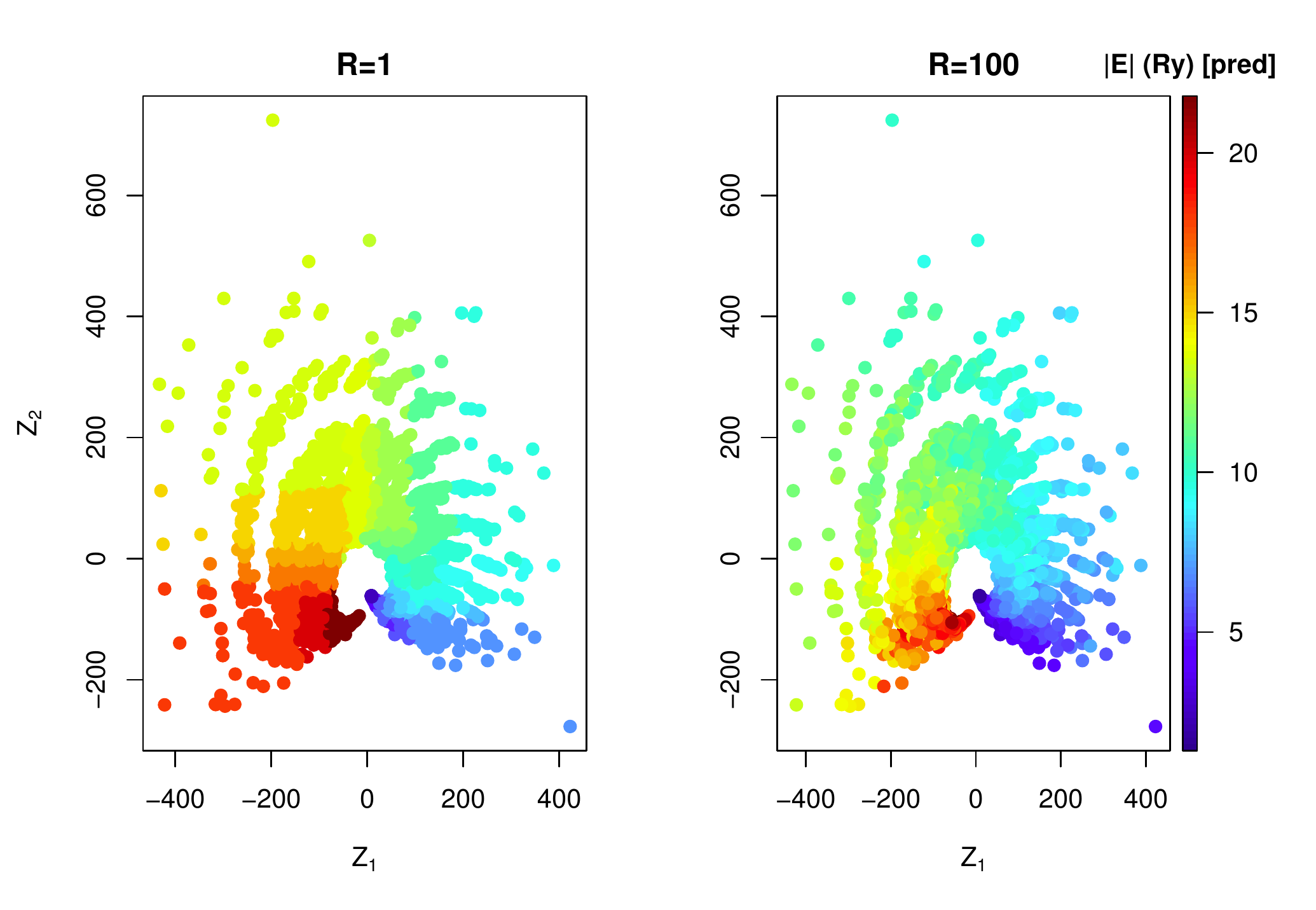}
\caption{\label{fig:xgbtree} Predicted $E_{\rm ps}$ using boosted regression trees for $R=1$ (left panel) and $R=100$ (right panel).
While a single tree ($R=1$) results in a rough division of the feature space, boosting many trees ($R=100$)
results in much higher accuracy.} 
\end{figure}

While the regression tree algorithm is simple and has low computational cost, it has low predictive accuracy. 
The tree shown in Fig.~\ref{fig:rpart} leads to a very rough description of the nonlinear behavior of E$_{\rm ps}$ as 
a function of the features, as can be seen in the regions of Fig.~\ref{fig:PCA}. An approach to more accurately describe the nonlinearities
is to train an ensemble of trees and combine the predictions, also known as boosting~\cite{Friedman:GBM}. 
The boosted tree algorithm starts with a null prediction (${\hat y}_i^{(0)} = 0$) and sequentially train trees 
on the residuals from the previous tree. Namely,
\begin{eqnarray}
{\hat y}_i^{(0)} &=& f_0(x_i) = 0, \,\,  r_i^{(0)} = y_i \nonumber\\
f_1 : r_i^{(0)},  \,\, {\hat y}_i^{(1)} &=& {\hat y}_i^{(0)} + \eta\, f_1, \,\, r_i^{(1)} = r_i^{(0)} - \eta\, f_1(x_i) \nonumber\\
&\vdots&  \nonumber\\
f_t : r_i^{(t-1)}, \,\, {\hat y}_i^{(t)} &=& {\hat y}_i^{(t-1)} + \eta\, f_t,  \,\, r_i^{(t)} = r_i^{(t-1)} - \eta\, f_t(x_i) \nonumber\\ 
& \vdots & \nonumber\\
f_{R} : r_i^{(R-1)}, \,\, {\hat y}_i^{(R)} &=& {\hat y}_i^{(R-1)} + \eta\, f_R, \,\,  r_i^{(R)} = r_i^{(R-1)} - \eta\, f_R(x_i) \nonumber 
\end{eqnarray} 
with the final prediction 
\begin{equation}
f({\bf X})  = \sum_{t=1}^R\, \eta\, f_t({\bf X}) \label{eqn:boost}
\end{equation}
where $f_t : r_i^{(t-1)}$ denotes that the tree at iteration $t$ is trained on the residuals from the prediction of the $(t-1)^{\rm th}$ tree.
The parameter $\eta$ is known as shrinkage, which determines the weight of the predictions that are added on at each iteration.
Both the shrinkage and the number of sequentially trained trees $R$ are parameters to be determined by cross-validation.
Fig.~\ref{fig:xgbtree} illustrates the process of boosting trees with $\{Z_1,Z_2\}$ as features.
While a single tree divides the feature space into rough regions (left panel in Fig.~\ref{fig:xgbtree}), with 100 trees (right panel of Fig.~\ref{fig:xgbtree}),
the predicted E$_{\rm ps}$ values are almost indistinguishable to the eye from the original data (Fig.~\ref{fig:PCA}).

In this study, we use the computationally efficient and scalable implementation of the boosting algorithm XGBoost~\cite{xgboost}. 
Apart from the shrinkage ($\eta$), number of trees ($R$), and the regularization term $\gamma$ (Eqn.(\ref{eqn:losstree})),  
XGBoost has several other parameters which are optimized to achieve highly accurate predictions. 
The parameters we have chosen to determine using cross-validation, with their short description are listed in Table~\ref{tab:params}, 
while a full description can be found in Ref.~\onlinecite{xgboost}.
\begin{table}[!ht]
\caption{\label{tab:params} XGBoost parameters}
\begin{ruledtabular}
\begin{tabular}{cc}
Parameter & Description \\
\hline
R & Number of trees grown \\ \\
$\eta$ & Shrinkage \\ \\
$\gamma$ & Regularization term  \\ \\
MD & Maximum number of  \\
    & terminal nodes ($T$) in a tree \\ \\
CST & Subsample ratio of randomly chosen \\ 
    & features while training a tree \\ \\
MCW & Minimum number of data points \\ 
    & in a region of feature space in each tree 
\end{tabular}
\end{ruledtabular}
\end{table}
\subsection{Neural Networks}

While neural networks are more popular for applications in classification problems, 
they can also be used for regression~\cite{elsl}. As input, neural networks take 
the feature vectors and using an activation function, nonlinear features are created and
used for regression. The activation function connects the linear input through layers of neurons
to the nonlinear output. The parameters used for connecting the layers in a neural network are
obtained by optimizing the RSS (in regression setting) using the training data. A 
structure of a neural network with one (hidden) layer is shown in Fig.~\ref{fig:nnet}. 
Given the input vector of features $x_i$, a nonlinear output is obtained by the following equations
\begin{eqnarray}
&& {\bf a}^{(1)} \leftarrow \left[ \begin{array}{c} 1 \\ x_1 \\ \vdots \\ x_p \end{array} \right], \,\, 
   {\bf a}^{(2)} = g\left( \theta^{(1)} \cdot {\bf a}^{(1)}\right) \nonumber\\
&& {\bf a}^{(2)} \leftarrow \left[ \begin{array}{c} 1 \\ a^{(2)}_1 \\ \vdots \\ a^{(2)}_h \end{array} \right], \,\, 
   {\bf {\hat y}} = ( \theta^{(2)} )^T \cdot {\bf a}^{(2)}  \label{eqn:nnet}
\end{eqnarray}
In the first line, the vector ${\bf a}^{(1)}$ is constructed from the input features ${\bf x}$ by adding 1, which
accounts for the constant term in regression. Then, the input vector ${\bf a}^{(1)}$ is transformed into a 
derived feature vector ${\bf a}^{(2)}$ in the hidden layer (with size $h$) using the nonlinear activation function $g(x)$. 
The transformation is parameterized by the coefficients $\theta^{(1)}_{ij}$ which are elements of a $(p+1) \times h$ matrix. 
The output layer, which serve as the prediction vector ${\hat {\bf y}}$ ($N$ dimensional), results from a linear 
transformation via $\theta^{(2)}_j$ which is an $N$ dimensional vector. The parameters $\theta^{(1)}$ and $\theta^{(2)}$ are obtained 
by minimizing Eqn.(\ref{eqn:loss}) with the following regularization term:
\begin{figure}[!ht]
\includegraphics[width=0.5\textwidth]{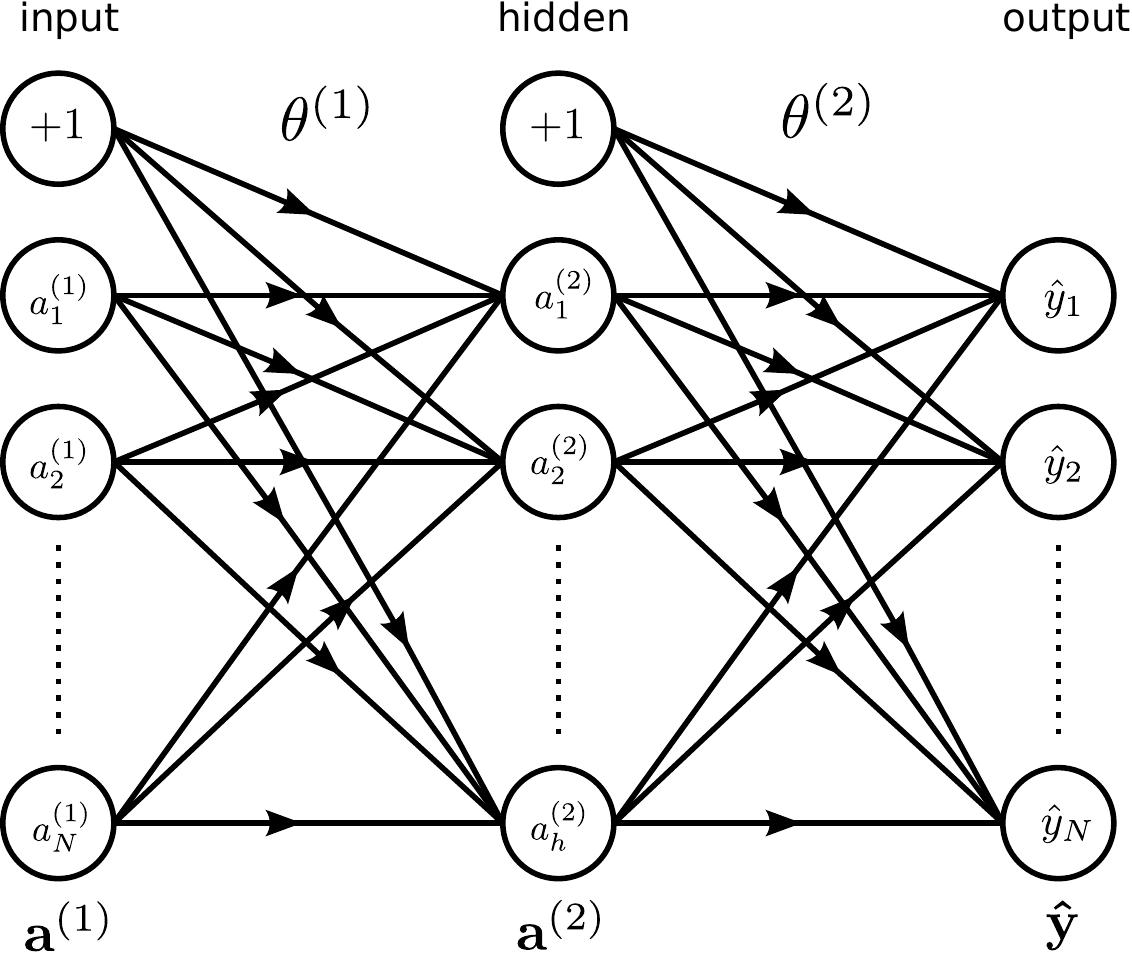}
\caption{\label{fig:nnet} A schematic representation of single layer neural network. Each line represents a term 
that connects a feature (${\bf a}^{(1)}$) to a derived feature (${\bf a}^{(2)}$) or a derived feature (${\bf a}^{(2)}$) 
to the output (${\hat {\bf y}}$). The activation functions use the parameters $\theta^{(1)}$ between the input and hidden layer,
while a linear transformation using $\theta^{(2)}$is used between the hidden layer and output.}
\end{figure}
\begin{equation}
R = \gamma\, \left[ \sum_{i=1}^{p}\, \sum_{k=1}^{h} (\theta^{(1)}_{ik})^2 + \sum_{k=1}^h\, (\theta^{(2)}_k)^2 \right] 
\end{equation}
There are several choices for the activation function, but we adopt here the most widely used sigmoid function which is
given by
\begin{equation}
g(\theta \cdot {\bf x}) = \frac{1}{e^{-\theta \cdot {\bf x}} + 1} \label{eqn:sigmoid}
\end{equation}
It is possible to obtain more complex nonlinear relationships between the features ${\bf x}$ and the output layer,
by including more hidden layers. While more hidden layers may result in higher accuracy, because of the 
added computational cost, we have limited our study to single-hidden-layer neural networks. In fact, 
the computational cost of training even a single-layer neural network is much higher than a boosted regression tree,
and its discussion is included for the sake of assessing the prediction accuracy of the latter method.

\section{\label{sec:res} Results and Discussion}

\subsection{Model training and Accuracy}
\begin{figure}[!ht]
\includegraphics[width=0.8\textwidth]{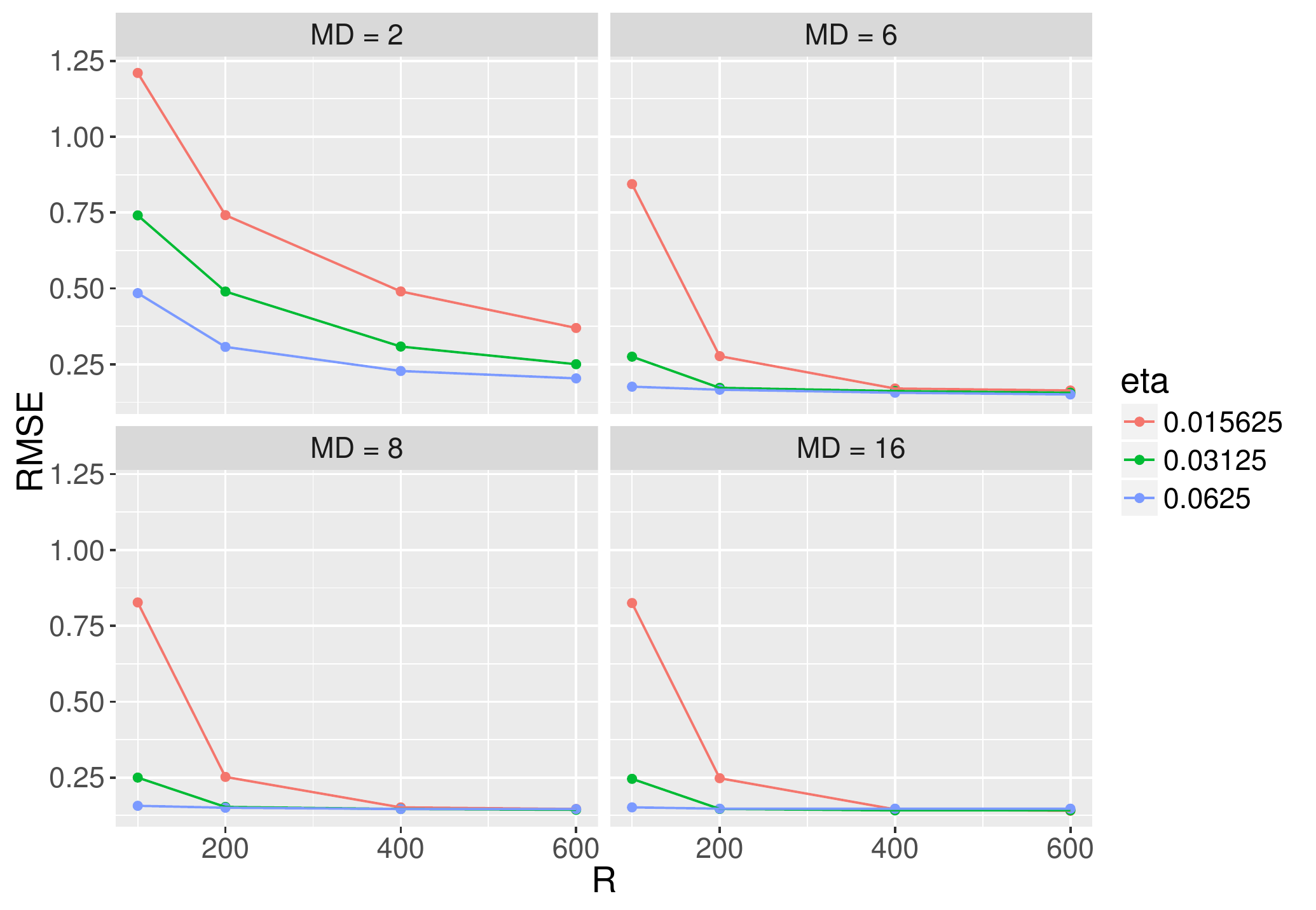}
\caption{\label{fig:cv5} RMSE (in Ry) from 5-fold cross-validation for boosted regression
trees for $\lambda^{(i)}$ used as predictors as a function of number of trees (R) and maximum number of trees (MD). 
Some of the parameters are fixed in the plots: MCW = 10, CS = 0.4, $\gamma$ = 0.0. }
\end{figure}

Using the cross-validation techniques outlined in the previous section, we have trained both boosted regression trees and 
neural networks. Either the Coulomb matrices ($C_{IJ}^{(i)}$) or their eigenspectrum ($\lambda^{(i)}$) were used as features
for training. The training set, 70\% of the full dataset chosen randomly,
is used to determine the parameters of the models.
The parameters obtained for boosted regression trees
using 5-fold cross-validation is shown in Table.~\ref{tab:xgb}.
\begin{table}[!ht]
\caption{\label{tab:xgb} Parameters determined for boosted regression trees using 5-fold cross-validation for 
$C_{IJ}^{(i)}$ and $\lambda^{(i)}$ based features. Definitions of the parameters are given in Table.~\ref{tab:params}.}
\begin{ruledtabular}
\begin{tabular}{ccccccc}
 & R & $\eta$ & $\gamma$ & MD & CST & MCW \\
\hline
$\lambda^{(i)}$ & 600 & 0.0156 & 0.0 & 16 & 0.4 & 10 \\
$C_{IJ}^{(i)}$ & 400 & 0.0625 & 0.0 & 6 & 0.2 & 10
\end{tabular}
\end{ruledtabular}
\end{table}
As an illustration, we show in Fig.~\ref{fig:cv5} the results of the 5-fold cross-validation with $\lambda^{(i)}$ as features.
Since the space of parameters to be optimized is multi-dimensional (Table.~\ref{tab:params}), we present
root mean squared error (RMSE) (i.e. the square root of MSE in Eqn.(\ref{eqn:MSE})), when MCW, CS and $\gamma$ are fixed to their 
optimized values from Table.~\ref{tab:xgb}.

\begin{figure}[!ht]
\includegraphics[width=0.6\textwidth]{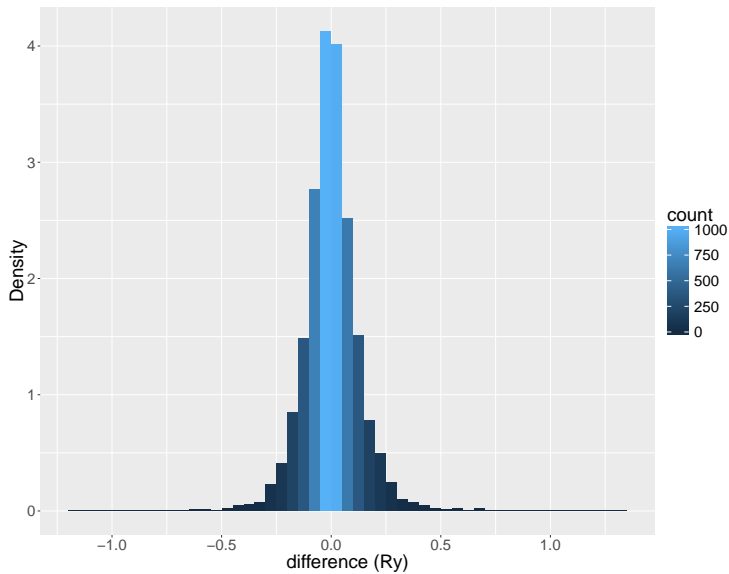}
\caption{\label{fig:act_vs_pred}
The difference between actual and predicted E$_{\rm ps}$ on the test set. Predictions are made
using the boosted regression tree with $\lambda^{(i)}$ as features. Color coding represents the number 
of molecules that fall into each bin in the histogram.}
\end{figure}

For neural networks, a 5-fold cross-validation approach is used for training the model when 
$\lambda^{(i)}$ are used as features, while the validation set approach is used when $C_{IJ}^{(i)}$ are used as
features due to heavy computational cost in this case. In the validation set approach, the initial training set (70\% of the 
original dataset) is split further into a validation set (40\% of initial training set) and a second level training set 
(60\% of initial training set) randomly. The cross-validation results in $h=25$ (size of the hidden layer) and
$\gamma = 1.0$ (the regularization parameter) when $\lambda^{(i)}$ are used, while $h=25$ and $\gamma = 0.1$
is obtained when $C_{IJ}^{(i)}$ are used.
The resulting cross-validation (or validation set) and test errors, measured by RMSE, for the trained models are summarized in Table.~\ref{tab:results}.
The best performance is obtained with the boosted regression tree algorithm when the  Coulomb matrices $C_{IJ}^{(i)}$ are used as features. 
The use of the eigenspectrum $\lambda^{(i)}$ as features only slightly increase the RMSE, while the computational cost
is much smaller. Therefore, it is reasonable to use the $\lambda^{(i)}$ instead of the full $C_{IJ}^{(i)}$.  
In Fig.~\ref{fig:act_vs_pred} we show the difference between predicted and actual E$_{\rm ps}$ evaluated on the test set,
using $\lambda^{(i)}$.

As we pointed out earlier, the boosted regression tree algorithm is computationally more efficient than neural networks. 
For example, with $\lambda^{(i)}$ as features, the elapsed time to train the boosted regression tree model (the model determined by
cross-validation whose parameters are listed in Table.~\ref{tab:xgb}) is 12.1 seconds (on a laptop with 8 CPU cores). Instead,
the elapsed time to train the neural network model (the model with $h=25$ and $\gamma=1.0$ determined by cross-validation) is 258.2 seconds.
In case of the elapsed time to obtain the fit to the testing data (after the models are trained), the neural network model takes 
0.02 seconds, while the boosted regression tree
model takes 0.01 seconds. In case of $C_{IJ}^{(i)}$ as features, the training times are 17.3 seconds for the boosted regression tree
and 16917.2 seconds for the neural network, while the fitting times to the test data are 0.01 and 0.2 seconds, respectively.
While the fitting time for neural networks are almost the same as boosted regression trees, the vast difference between 
training times, in addition to increased accuracy, makes boosted regression tree the preferred algorithm.

Previous studies which analyzed the GDB database~\cite{fink:gdb13-1,fink:gdb13-2} of molecules, have found
much smaller test RMSE for their best achieving learning methods~\cite{Rupp:jctc,Rupp:NJP}. 
For example, Ref.~\onlinecite{Rupp:jctc} reported a RMSE of 20.29 kcal/mol for multi-layer neural networks
and 13.18 kcal/mol for a nonlinear kernel method when $\lambda^{(i)}$ were used as features. These are smaller values of RMSE compared to 
our values of 41.81 and 60.06 kcal/mol for boosted trees and single layer neural networks, respectively.
The main difference is due to the fact that the variability in the GDB database is much smaller
than our dataset which is based on the PubChem data. 
Ref.~\onlinecite{Rupp:jctc} used a subset of the GDB database that contains 7165 molecular structures which has maximum 23 atoms 
per molecule. The reported standard deviation of the atomization energies 
(RMSE when mean value of ${\bf y}$ is used as the prediction) was 223.92 kcal/mol. 
In our study, we have 16,242 structures with a maximum of 50 atoms per molecule, while the standard deviation 
of the atomization energies is 1147.89 kcal/mol, which indicates that the dataset used in this study
encompasses a larger range of molecules. 
As a result of the larger variance in the training data (almost five times of Ref.~\onlinecite{Rupp:jctc}), 
the learning algorithms result in higher RMSE values. Notice also that the number of molecules in this study is 
much larger, leading to a computationally more expensive model training when a method like neural network regression is used.
It is possible to obtain much better accuracies by including more data with the use of the full PubChem database, instead
of the first 75,000 entries as we did in this work. In addition, using several copies of the same molecule by representing them via randomly re-ordered Coulomb matrices 
(a method introduced in Ref.~\onlinecite{Rupp:jctc} to address the uniqueness problem of $C_{IJ}^{(i)}$) 
would reduce the variance in the dataset, leading to better accuracies.
While the reported RMSE values are higher than what would be desired (e.g. the accuracy of a few kcal/mol), inclusion of more data
presents a clear path to reach more accurate models. As an example, we have tested the accuracy of the randomly re-ordered 
Coulomb matrices and found that with only 4 random re-orderings included (yielding a training set four times larger than original),
the RMSE of 36.63 kcal/mol (Table.~\ref{tab:results}) reduces to 27.74 kcal/mol.
It is also possible to use an ensemble method~\cite{Friedman:Ensemble}, where predictions of boosted trees and neural networks are combined. In each method,
the worst predictions are on different molecules, and by combining the two on the regions where they perform the best, 
improved accuracies can be obtained. 
Another approach has been proposed in a more recent work, where feature learning increased prediction
accuracies~\cite{Dai:latent}.
While all of these approaches would lead to increased accuracy, they also come with added computational cost and are
beyond the scope of this work. Therefore, we leave these tasks for a future study.

\begin{table}
\caption{\label{tab:results} 
5-fold CV (validation set in case of neural network with $C_{IJ}^{(i)}$) and test RMSE in kcal/mol. 
The difference between 5-fold CV and test RMSE is smaller than that of validation set and test RMSE as expected,
since 5-fold CV reduces the dependence of RMSE to the way the data is split.}
\begin{ruledtabular}
\begin{tabular}{cccc}
Feature & Method  & 5-fold CV (or validation set) error & test error \\
\hline
$\lambda^{(i)}$ & Boosted Tree & 44.09 & 41.81 \\
                & N. Network   & 62.15 & 60.06 \\
\hline
$C_{IJ}^{(i)}$  & Boosted Tree & 38.75 & 36.63 \\
                & N. Network   & 45.25 & 58.39 
\end{tabular}
\end{ruledtabular}
\end{table}

\subsection{Predictions on independent datasets}
In the previous subsection, we have trained models to predict E$_{\rm ps}$ based on 
a dataset of molecules made up of elements in the CHNOPS set. Namely, each molecule contains
at least one of the atoms in the set 
$\{{\rm C}, {\rm H}, {\rm N}, {\rm O}, {\rm P}, {\rm S} \}$. 
While a random division of the dataset into training and test
sets allowed us to evaluate the accuracy of the models, another possible measure for performance would be
based on a set of molecules outside the CHNOPS set. For this purpose, we have constructed two distinct datasets
in order to test the models trained in the previous subsection: (i) Cl-set, which contains at least one Cl atom
in addition to at least one of CHNOPS, (ii) Si-set, which contains at least one Si atom in addition to at least 
one of CHNOPS. From the first 75,000 entries in the PubChem database, we have found that there are
159 molecules in the Si-set and 4114 in the Cl-set. 
Similar to the original CHNOPS set, each molecule in both Si and Cl sets have a maximum of 
50 atoms per molecule, and even number of electrons.
Using the boosted regression trees we have trained in the previous subsection, we 
predict E$_{\rm ps}$ in Cl and Si sets, for which the results are 
presented in Table.~\ref{tab:ClSi}. Since the models are not trained with molecules comprising of Si or Cl, 
this test assesses applicability of our method when predictions on molecules with new elements need to 
be obtained.
\begin{table}
\caption{\label{tab:ClSi} Test RMSE (kcal/mol) on the Cl and Si sets. Predictions are based on the boosted regression tree
algorithm trained on the CHNOPS set.}
\begin{ruledtabular}
\begin{tabular}{ccc}
 Set & RMSE ($\lambda^{(i)}$) & RMSE ($C_{IJ}^{(i)}$) \\
 \hline
 Cl & 75.18 & 114.20 \\
 Si & 89.13 & 86.53 
\end{tabular}
\end{ruledtabular}
\end{table}
As expected, the RMSE values are higher for the Cl and Si sets, than that of the test errors reported in Table.~\ref{tab:results}.
Therefore the models can only be used as exploratory tools
when predictions on completely separate datasets are needed.

\section{\label{sec:conc} Conclusions}
In this work, we have proposed the use of boosted regression trees as a higher accuracy and computationally much more
efficient alternative to some other machine learning methods proposed for electronic structure prediction. 
We have tested the performance of boosted regression trees using the PubChem database, and shown that
it outperforms single-layer neural networks in predicting ground state energies (equivalently E$_{\rm ps}$).
Due to the ability to grow many trees in parallel, boosted regression trees are much faster than neural networks, 
which require large matrix operations. We have also shown that the trained algorithms can be used for predicting
electronic structure of molecules containing elements other than the ones included in the training set. While
the prediction accuracy is reduced in this case, the method is still applicable for exploratory studies.

Machine learning techniques provide a rich possibility of applications for quantum mechanical simulations in 
computational chemistry and materials science~\cite{Rupp:Rev}. In fact, there has been several other compelling applications
of learning algorithms for predicting electronic properties other than atomization energies for molecules~\cite{Rupp:NJP},
prediction of electronic structure of solids~\cite{Gross:Solids,Faber:Solids}, finding DFT functionals~\cite{Burke:ML}
and determining potential energy surfaces~\cite{PES-1,PES-2,Khorshidi}. 
The adoption of boosted regression trees for the learning method, as proposed here, would reduce the cost of 
model training compared to computationally heavier algorithms like neural networks and support vector regression,
without sacrificing, and possibly increasing prediction accuracies.

With the ability to predict electronic properties without performing
new simulations for each molecule, machine learning techniques open up exciting pathways for rational design of new compounds.
Combined with numerous efforts to catalog and standardize datasets, these methods will 
be invaluable for many scientific and technological applications.

\section{Acknowledgements}
We acknowledge support from the Center for Scientific Computing from the CNSI, MRL: an NSF MRSEC (DMR-1121053) and NSF CNS-0960316.

\bibliography{chnops}

\providecommand{\noopsort}[1]{}\providecommand{\singleletter}[1]{#1}%
\begin{thebibliography}{37}%
\makeatletter
\providecommand \@ifxundefined [1]{%
 \@ifx{#1\undefined}
}%
\providecommand \@ifnum [1]{%
 \ifnum #1\expandafter \@firstoftwo
 \else \expandafter \@secondoftwo
 \fi
}%
\providecommand \@ifx [1]{%
 \ifx #1\expandafter \@firstoftwo
 \else \expandafter \@secondoftwo
 \fi
}%
\providecommand \natexlab [1]{#1}%
\providecommand \enquote  [1]{``#1''}%
\providecommand \bibnamefont  [1]{#1}%
\providecommand \bibfnamefont [1]{#1}%
\providecommand \citenamefont [1]{#1}%
\providecommand \href@noop [0]{\@secondoftwo}%
\providecommand \href [0]{\begingroup \@sanitize@url \@href}%
\providecommand \@href[1]{\@@startlink{#1}\@@href}%
\providecommand \@@href[1]{\endgroup#1\@@endlink}%
\providecommand \@sanitize@url [0]{\catcode `\\12\catcode `\$12\catcode
  `\&12\catcode `\#12\catcode `\^12\catcode `\_12\catcode `\%12\relax}%
\providecommand \@@startlink[1]{}%
\providecommand \@@endlink[0]{}%
\providecommand \url  [0]{\begingroup\@sanitize@url \@url }%
\providecommand \@url [1]{\endgroup\@href {#1}{\urlprefix }}%
\providecommand \urlprefix  [0]{URL }%
\providecommand \Eprint [0]{\href }%
\providecommand \doibase [0]{http://dx.doi.org/}%
\providecommand \selectlanguage [0]{\@gobble}%
\providecommand \bibinfo  [0]{\@secondoftwo}%
\providecommand \bibfield  [0]{\@secondoftwo}%
\providecommand \translation [1]{[#1]}%
\providecommand \BibitemOpen [0]{}%
\providecommand \bibitemStop [0]{}%
\providecommand \bibitemNoStop [0]{.\EOS\space}%
\providecommand \EOS [0]{\spacefactor3000\relax}%
\providecommand \BibitemShut  [1]{\csname bibitem#1\endcsname}%
\let\auto@bib@innerbib\@empty
\bibitem [{\citenamefont {Jain}\ \emph {et~al.}(2013)\citenamefont {Jain},
  \citenamefont {Ong}, \citenamefont {Hautier}, \citenamefont {Chen},
  \citenamefont {Richards}, \citenamefont {Dacek}, \citenamefont {Cholia},
  \citenamefont {Gunter}, \citenamefont {Skinner}, \citenamefont {Ceder},\ and\
  \citenamefont {Persson}}]{Jain:MGenome}%
  \BibitemOpen
  \bibfield  {author} {\bibinfo {author} {\bibfnamefont {A.}~\bibnamefont
  {Jain}}, \bibinfo {author} {\bibfnamefont {S.~P.}\ \bibnamefont {Ong}},
  \bibinfo {author} {\bibfnamefont {G.}~\bibnamefont {Hautier}}, \bibinfo
  {author} {\bibfnamefont {W.}~\bibnamefont {Chen}}, \bibinfo {author}
  {\bibfnamefont {W.~D.}\ \bibnamefont {Richards}}, \bibinfo {author}
  {\bibfnamefont {S.}~\bibnamefont {Dacek}}, \bibinfo {author} {\bibfnamefont
  {S.}~\bibnamefont {Cholia}}, \bibinfo {author} {\bibfnamefont
  {D.}~\bibnamefont {Gunter}}, \bibinfo {author} {\bibfnamefont
  {D.}~\bibnamefont {Skinner}}, \bibinfo {author} {\bibfnamefont
  {G.}~\bibnamefont {Ceder}}, \ and\ \bibinfo {author} {\bibfnamefont {K.~A.}\
  \bibnamefont {Persson}},\ }\href@noop {} {\bibfield  {journal} {\bibinfo
  {journal} {APL Mater.}\ }\textbf {\bibinfo {volume} {1}},\ \bibinfo {pages}
  {011002} (\bibinfo {year} {2013})}\BibitemShut {NoStop}%
\bibitem [{\citenamefont {Curtarolo}\ \emph {et~al.}(2012)\citenamefont
  {Curtarolo}, \citenamefont {Setyawan}, \citenamefont {Hart}, \citenamefont
  {M}, \citenamefont {Chepulskii}, \citenamefont {Taylor}, \citenamefont
  {Wang}, \citenamefont {Xue}, \citenamefont {Yang}, \citenamefont {O.},
  \citenamefont {Mehl}, \citenamefont {Stokes}, \citenamefont {Demchenko},\
  and\ \citenamefont {Morgan}}]{Curtarolo:AFLOW}%
  \BibitemOpen
  \bibfield  {author} {\bibinfo {author} {\bibfnamefont {S.}~\bibnamefont
  {Curtarolo}}, \bibinfo {author} {\bibfnamefont {W.}~\bibnamefont {Setyawan}},
  \bibinfo {author} {\bibfnamefont {G.~L.~W.}\ \bibnamefont {Hart}}, \bibinfo
  {author} {\bibfnamefont {J.}~\bibnamefont {M}}, \bibinfo {author}
  {\bibfnamefont {R.~V.}\ \bibnamefont {Chepulskii}}, \bibinfo {author}
  {\bibfnamefont {R.~H.}\ \bibnamefont {Taylor}}, \bibinfo {author}
  {\bibfnamefont {S.}~\bibnamefont {Wang}}, \bibinfo {author} {\bibfnamefont
  {J.}~\bibnamefont {Xue}}, \bibinfo {author} {\bibfnamefont {K.}~\bibnamefont
  {Yang}}, \bibinfo {author} {\bibfnamefont {L.}~\bibnamefont {O.}}, \bibinfo
  {author} {\bibfnamefont {M.~J.}\ \bibnamefont {Mehl}}, \bibinfo {author}
  {\bibfnamefont {H.~T.}\ \bibnamefont {Stokes}}, \bibinfo {author}
  {\bibfnamefont {D.~O.}\ \bibnamefont {Demchenko}}, \ and\ \bibinfo {author}
  {\bibfnamefont {D.}~\bibnamefont {Morgan}},\ }\href@noop {} {\bibfield
  {journal} {\bibinfo  {journal} {Comput. Mater. Sci.}\ }\textbf {\bibinfo
  {volume} {58}},\ \bibinfo {pages} {218} (\bibinfo {year} {2012})}\BibitemShut
  {NoStop}%
\bibitem [{\citenamefont {Hart}\ \emph {et~al.}(2013)\citenamefont {Hart},
  \citenamefont {Curtarolo}, \citenamefont {Massalski},\ and\ \citenamefont
  {Levy}}]{Curtarolo:Alloys}%
  \BibitemOpen
  \bibfield  {author} {\bibinfo {author} {\bibfnamefont {G.~L.~W.}\
  \bibnamefont {Hart}}, \bibinfo {author} {\bibfnamefont {S.}~\bibnamefont
  {Curtarolo}}, \bibinfo {author} {\bibfnamefont {T.~B.}\ \bibnamefont
  {Massalski}}, \ and\ \bibinfo {author} {\bibfnamefont {O.}~\bibnamefont
  {Levy}},\ }\href@noop {} {\bibfield  {journal} {\bibinfo  {journal} {Phys.
  Rev. X}\ }\textbf {\bibinfo {volume} {3}},\ \bibinfo {pages} {041035}
  (\bibinfo {year} {2013})}\BibitemShut {NoStop}%
\bibitem [{\citenamefont {Hautier}\ \emph {et~al.}(2010)\citenamefont
  {Hautier}, \citenamefont {Fischer}, \citenamefont {Jain}, \citenamefont
  {Mueller},\ and\ \citenamefont {Ceder}}]{Hautier:Oxide}%
  \BibitemOpen
  \bibfield  {author} {\bibinfo {author} {\bibfnamefont {G.}~\bibnamefont
  {Hautier}}, \bibinfo {author} {\bibfnamefont {C.~C.}\ \bibnamefont
  {Fischer}}, \bibinfo {author} {\bibfnamefont {A.}~\bibnamefont {Jain}},
  \bibinfo {author} {\bibfnamefont {T.}~\bibnamefont {Mueller}}, \ and\
  \bibinfo {author} {\bibfnamefont {G.}~\bibnamefont {Ceder}},\ }\href@noop {}
  {\bibfield  {journal} {\bibinfo  {journal} {Chem. Mater.}\ }\textbf {\bibinfo
  {volume} {22}},\ \bibinfo {pages} {3762} (\bibinfo {year}
  {2010})}\BibitemShut {NoStop}%
\bibitem [{\citenamefont {Pizzi}\ \emph {et~al.}(2016)\citenamefont {Pizzi},
  \citenamefont {Cepellotti}, \citenamefont {Sabatini}, \citenamefont
  {Marzari},\ and\ \citenamefont {Kozinsky}}]{Pizzi:Aiida}%
  \BibitemOpen
  \bibfield  {author} {\bibinfo {author} {\bibfnamefont {G.}~\bibnamefont
  {Pizzi}}, \bibinfo {author} {\bibfnamefont {A.}~\bibnamefont {Cepellotti}},
  \bibinfo {author} {\bibfnamefont {R.}~\bibnamefont {Sabatini}}, \bibinfo
  {author} {\bibfnamefont {N.}~\bibnamefont {Marzari}}, \ and\ \bibinfo
  {author} {\bibfnamefont {B.}~\bibnamefont {Kozinsky}},\ }\href@noop {}
  {\bibfield  {journal} {\bibinfo  {journal} {Comput. Mater. Sci.}\ }\textbf
  {\bibinfo {volume} {111}},\ \bibinfo {pages} {218} (\bibinfo {year}
  {2016})}\BibitemShut {NoStop}%
\bibitem [{\citenamefont {Hohenberg}\ and\ \citenamefont {Kohn}(1964)}]{DFT-1}%
  \BibitemOpen
  \bibfield  {author} {\bibinfo {author} {\bibfnamefont {P.}~\bibnamefont
  {Hohenberg}}\ and\ \bibinfo {author} {\bibfnamefont {W.}~\bibnamefont
  {Kohn}},\ }\href@noop {} {\bibfield  {journal} {\bibinfo  {journal} {Phys.
  Rev.}\ }\textbf {\bibinfo {volume} {136}},\ \bibinfo {pages} {B864} (\bibinfo
  {year} {1964})}\BibitemShut {NoStop}%
\bibitem [{\citenamefont {Kohn}\ and\ \citenamefont {Sham}(1965)}]{DFT-2}%
  \BibitemOpen
  \bibfield  {author} {\bibinfo {author} {\bibfnamefont {W.}~\bibnamefont
  {Kohn}}\ and\ \bibinfo {author} {\bibfnamefont {L.~J.}\ \bibnamefont
  {Sham}},\ }\href@noop {} {\bibfield  {journal} {\bibinfo  {journal} {Phys.
  Rev.}\ }\textbf {\bibinfo {volume} {140}},\ \bibinfo {pages} {A1133}
  (\bibinfo {year} {1965})}\BibitemShut {NoStop}%
\bibitem [{\citenamefont {Behler}(2011)}]{Behler:nnet}%
  \BibitemOpen
  \bibfield  {author} {\bibinfo {author} {\bibfnamefont {J.}~\bibnamefont
  {Behler}},\ }\href@noop {} {\bibfield  {journal} {\bibinfo  {journal} {Phys.
  Chem. Chem. Phys.}\ }\textbf {\bibinfo {volume} {13}},\ \bibinfo {pages}
  {17930} (\bibinfo {year} {2011})}\BibitemShut {NoStop}%
\bibitem [{\citenamefont {Hu}\ \emph {et~al.}(2003)\citenamefont {Hu},
  \citenamefont {Wang}, \citenamefont {Wong},\ and\ \citenamefont
  {Chen}}]{Hu:nnet}%
  \BibitemOpen
  \bibfield  {author} {\bibinfo {author} {\bibfnamefont {L.}~\bibnamefont
  {Hu}}, \bibinfo {author} {\bibfnamefont {X.}~\bibnamefont {Wang}}, \bibinfo
  {author} {\bibfnamefont {L.}~\bibnamefont {Wong}}, \ and\ \bibinfo {author}
  {\bibfnamefont {G.}~\bibnamefont {Chen}},\ }\href@noop {} {\bibfield
  {journal} {\bibinfo  {journal} {J. Chem. Phys.}\ }\textbf {\bibinfo {volume}
  {119}},\ \bibinfo {pages} {11501} (\bibinfo {year} {2003})}\BibitemShut
  {NoStop}%
\bibitem [{\citenamefont {Balabin}\ and\ \citenamefont
  {Lomakina}(2009)}]{Balabin:nnet}%
  \BibitemOpen
  \bibfield  {author} {\bibinfo {author} {\bibfnamefont {R.~M.}\ \bibnamefont
  {Balabin}}\ and\ \bibinfo {author} {\bibfnamefont {E.~I.}\ \bibnamefont
  {Lomakina}},\ }\href@noop {} {\bibfield  {journal} {\bibinfo  {journal} {J.
  Chem. Phys.}\ }\textbf {\bibinfo {volume} {131}},\ \bibinfo {pages} {074104}
  (\bibinfo {year} {2009})}\BibitemShut {NoStop}%
\bibitem [{\citenamefont {Sun}\ \emph {et~al.}(2014)\citenamefont {Sun},
  \citenamefont {Wu}, \citenamefont {Song}, \citenamefont {Hu}, \citenamefont
  {Shan},\ and\ \citenamefont {Chen}}]{Sun:nnet}%
  \BibitemOpen
  \bibfield  {author} {\bibinfo {author} {\bibfnamefont {J.}~\bibnamefont
  {Sun}}, \bibinfo {author} {\bibfnamefont {J.}~\bibnamefont {Wu}}, \bibinfo
  {author} {\bibfnamefont {T.}~\bibnamefont {Song}}, \bibinfo {author}
  {\bibfnamefont {L.}~\bibnamefont {Hu}}, \bibinfo {author} {\bibfnamefont
  {K.}~\bibnamefont {Shan}}, \ and\ \bibinfo {author} {\bibfnamefont
  {G.}~\bibnamefont {Chen}},\ }\href@noop {} {\bibfield  {journal} {\bibinfo
  {journal} {J. Phys. Chem. A}\ }\textbf {\bibinfo {volume} {118}},\ \bibinfo
  {pages} {9120} (\bibinfo {year} {2014})}\BibitemShut {NoStop}%
\bibitem [{\citenamefont {Balabin}\ and\ \citenamefont
  {Lomakina}(2011)}]{Balabin:svm}%
  \BibitemOpen
  \bibfield  {author} {\bibinfo {author} {\bibfnamefont {R.~M.}\ \bibnamefont
  {Balabin}}\ and\ \bibinfo {author} {\bibfnamefont {E.~I.}\ \bibnamefont
  {Lomakina}},\ }\href@noop {} {\bibfield  {journal} {\bibinfo  {journal}
  {Phys. Chem. Chem. Phys.}\ }\textbf {\bibinfo {volume} {13}},\ \bibinfo
  {pages} {11710} (\bibinfo {year} {2011})}\BibitemShut {NoStop}%
\bibitem [{\citenamefont {Lee}\ \emph {et~al.}(2016)\citenamefont {Lee},
  \citenamefont {Seko}, \citenamefont {Shitara}, \citenamefont {Nakayama},\
  and\ \citenamefont {Tanaka}}]{Lee:svr}%
  \BibitemOpen
  \bibfield  {author} {\bibinfo {author} {\bibfnamefont {J.}~\bibnamefont
  {Lee}}, \bibinfo {author} {\bibfnamefont {A.}~\bibnamefont {Seko}}, \bibinfo
  {author} {\bibfnamefont {K.}~\bibnamefont {Shitara}}, \bibinfo {author}
  {\bibfnamefont {K.}~\bibnamefont {Nakayama}}, \ and\ \bibinfo {author}
  {\bibfnamefont {I.}~\bibnamefont {Tanaka}},\ }\href@noop {} {\bibfield
  {journal} {\bibinfo  {journal} {Phys. Rev. B}\ }\textbf {\bibinfo {volume}
  {93}},\ \bibinfo {pages} {115104} (\bibinfo {year} {2016})}\BibitemShut
  {NoStop}%
\bibitem [{Note1()}]{Note1}%
  \BibitemOpen
  \bibinfo {note} {In practice, the situation is complicated further by the
  choice of pseudopotentials, basis sets, kinetic energy cut-offs, boundary
  conditions (especially for molecular systems) and the simulation code~\cite
  {reproducibility}}\BibitemShut {NoStop}%
\bibitem [{\citenamefont {Rupp}\ \emph {et~al.}(2012)\citenamefont {Rupp},
  \citenamefont {Tkatchenko}, \citenamefont {M\"uller},\ and\ \citenamefont
  {von Lilienfeld}}]{Rupp:CIJ}%
  \BibitemOpen
  \bibfield  {author} {\bibinfo {author} {\bibfnamefont {M.}~\bibnamefont
  {Rupp}}, \bibinfo {author} {\bibfnamefont {A.}~\bibnamefont {Tkatchenko}},
  \bibinfo {author} {\bibfnamefont {K.-R.}\ \bibnamefont {M\"uller}}, \ and\
  \bibinfo {author} {\bibfnamefont {O.~A.}\ \bibnamefont {von Lilienfeld}},\
  }\href@noop {} {\bibfield  {journal} {\bibinfo  {journal} {Phys. Rev. Lett.}\
  }\textbf {\bibinfo {volume} {108}},\ \bibinfo {pages} {058301} (\bibinfo
  {year} {2012})}\BibitemShut {NoStop}%
\bibitem [{\citenamefont {Hansen}\ \emph {et~al.}(2013)\citenamefont {Hansen},
  \citenamefont {Montavon}, \citenamefont {Biegler}, \citenamefont {Fazli},
  \citenamefont {Rupp}, \citenamefont {Scheffler}, \citenamefont {von
  Lilienfeld}, \citenamefont {Tkatchenko},\ and\ \citenamefont
  {Müller}}]{Rupp:jctc}%
  \BibitemOpen
  \bibfield  {author} {\bibinfo {author} {\bibfnamefont {K.}~\bibnamefont
  {Hansen}}, \bibinfo {author} {\bibfnamefont {G.}~\bibnamefont {Montavon}},
  \bibinfo {author} {\bibfnamefont {F.}~\bibnamefont {Biegler}}, \bibinfo
  {author} {\bibfnamefont {S.}~\bibnamefont {Fazli}}, \bibinfo {author}
  {\bibfnamefont {M.}~\bibnamefont {Rupp}}, \bibinfo {author} {\bibfnamefont
  {M.}~\bibnamefont {Scheffler}}, \bibinfo {author} {\bibfnamefont {O.~A.}\
  \bibnamefont {von Lilienfeld}}, \bibinfo {author} {\bibfnamefont
  {A.}~\bibnamefont {Tkatchenko}}, \ and\ \bibinfo {author} {\bibfnamefont
  {K.-R.}\ \bibnamefont {Müller}},\ }\href@noop {} {\bibfield  {journal}
  {\bibinfo  {journal} {J. Chem. Theory Comput.}\ }\textbf {\bibinfo {volume}
  {9}},\ \bibinfo {pages} {3404} (\bibinfo {year} {2013})}\BibitemShut
  {NoStop}%
\bibitem [{\citenamefont {Hill}\ \emph {et~al.}(2016)\citenamefont {Hill},
  \citenamefont {Mulholland}, \citenamefont {Persson}, \citenamefont
  {Seshadri}, \citenamefont {Wolverton},\ and\ \citenamefont
  {Meredig}}]{Hill:MRS}%
  \BibitemOpen
  \bibfield  {author} {\bibinfo {author} {\bibfnamefont {J.}~\bibnamefont
  {Hill}}, \bibinfo {author} {\bibfnamefont {G.}~\bibnamefont {Mulholland}},
  \bibinfo {author} {\bibfnamefont {K.}~\bibnamefont {Persson}}, \bibinfo
  {author} {\bibfnamefont {R.}~\bibnamefont {Seshadri}}, \bibinfo {author}
  {\bibfnamefont {C.}~\bibnamefont {Wolverton}}, \ and\ \bibinfo {author}
  {\bibfnamefont {B.}~\bibnamefont {Meredig}},\ }\href@noop {} {\bibfield
  {journal} {\bibinfo  {journal} {MRS Bull.}\ }\textbf {\bibinfo {volume}
  {41}},\ \bibinfo {pages} {399} (\bibinfo {year} {2016})}\BibitemShut
  {NoStop}%
\bibitem [{\citenamefont {Kim}\ \emph {et~al.}(2015)\citenamefont {Kim},
  \citenamefont {Thiessen}, \citenamefont {Bolton}, \citenamefont {Chen},
  \citenamefont {Fu}, \citenamefont {Gindulyte}, \citenamefont {Han},
  \citenamefont {He}, \citenamefont {He}, \citenamefont {Shoemaker},
  \citenamefont {Wang}, \citenamefont {Yu}, \citenamefont {Zhang},\ and\
  \citenamefont {Bryant}}]{pubchem}%
  \BibitemOpen
  \bibfield  {author} {\bibinfo {author} {\bibfnamefont {S.}~\bibnamefont
  {Kim}}, \bibinfo {author} {\bibfnamefont {P.~A.}\ \bibnamefont {Thiessen}},
  \bibinfo {author} {\bibfnamefont {E.~E.}\ \bibnamefont {Bolton}}, \bibinfo
  {author} {\bibfnamefont {J.}~\bibnamefont {Chen}}, \bibinfo {author}
  {\bibfnamefont {G.}~\bibnamefont {Fu}}, \bibinfo {author} {\bibfnamefont
  {A.}~\bibnamefont {Gindulyte}}, \bibinfo {author} {\bibfnamefont
  {L.}~\bibnamefont {Han}}, \bibinfo {author} {\bibfnamefont {J.}~\bibnamefont
  {He}}, \bibinfo {author} {\bibfnamefont {S.}~\bibnamefont {He}}, \bibinfo
  {author} {\bibfnamefont {B.~A.}\ \bibnamefont {Shoemaker}}, \bibinfo {author}
  {\bibfnamefont {J.}~\bibnamefont {Wang}}, \bibinfo {author} {\bibfnamefont
  {B.}~\bibnamefont {Yu}}, \bibinfo {author} {\bibfnamefont {J.}~\bibnamefont
  {Zhang}}, \ and\ \bibinfo {author} {\bibfnamefont {S.~H.}\ \bibnamefont
  {Bryant}},\ }\href@noop {} {\bibfield  {journal} {\bibinfo  {journal}
  {Nucleic Acids Res.}\ ,\ \bibinfo {pages} {gkv951}} (\bibinfo {year}
  {2015})}\BibitemShut {NoStop}%
\bibitem [{\citenamefont {Friedman}(2001)}]{Friedman:GBM}%
  \BibitemOpen
  \bibfield  {author} {\bibinfo {author} {\bibfnamefont {J.~H.}\ \bibnamefont
  {Friedman}},\ }\href@noop {} {\bibfield  {journal} {\bibinfo  {journal} {Ann.
  Statist.}\ }\textbf {\bibinfo {volume} {29}},\ \bibinfo {pages} {1189}
  (\bibinfo {year} {2001})}\BibitemShut {NoStop}%
\bibitem [{\citenamefont {Giannozzi}\ \emph {et~al.}(2009)\citenamefont
  {Giannozzi}, \citenamefont {Baroni}, \citenamefont {Bonini}, \citenamefont
  {Calandra}, \citenamefont {Car}, \citenamefont {Cavazzoni}, \citenamefont
  {Ceresoli}, \citenamefont {Chiarotti}, \citenamefont {Cococcioni},
  \citenamefont {Dabo} \emph {et~al.}}]{QE}%
  \BibitemOpen
  \bibfield  {author} {\bibinfo {author} {\bibfnamefont {P.}~\bibnamefont
  {Giannozzi}}, \bibinfo {author} {\bibfnamefont {S.}~\bibnamefont {Baroni}},
  \bibinfo {author} {\bibfnamefont {N.}~\bibnamefont {Bonini}}, \bibinfo
  {author} {\bibfnamefont {M.}~\bibnamefont {Calandra}}, \bibinfo {author}
  {\bibfnamefont {R.}~\bibnamefont {Car}}, \bibinfo {author} {\bibfnamefont
  {C.}~\bibnamefont {Cavazzoni}}, \bibinfo {author} {\bibfnamefont
  {D.}~\bibnamefont {Ceresoli}}, \bibinfo {author} {\bibfnamefont
  {G.}~\bibnamefont {Chiarotti}}, \bibinfo {author} {\bibfnamefont
  {M.}~\bibnamefont {Cococcioni}}, \bibinfo {author} {\bibfnamefont
  {I.}~\bibnamefont {Dabo}},  \emph {et~al.},\ }\href@noop {} {\bibfield
  {journal} {\bibinfo  {journal} {J. Phys. Condens. Matter}\ }\textbf {\bibinfo
  {volume} {21}},\ \bibinfo {pages} {395502} (\bibinfo {year}
  {2009})}\BibitemShut {NoStop}%
\bibitem [{\citenamefont {Perdew}\ \emph {et~al.}(1996)\citenamefont {Perdew},
  \citenamefont {Burke},\ and\ \citenamefont {Ernzerhof}}]{pbe}%
  \BibitemOpen
  \bibfield  {author} {\bibinfo {author} {\bibfnamefont {J.~P.}\ \bibnamefont
  {Perdew}}, \bibinfo {author} {\bibfnamefont {K.}~\bibnamefont {Burke}}, \
  and\ \bibinfo {author} {\bibfnamefont {M.}~\bibnamefont {Ernzerhof}},\
  }\href@noop {} {\bibfield  {journal} {\bibinfo  {journal} {Phys. Rev. Lett.}\
  }\textbf {\bibinfo {volume} {77}},\ \bibinfo {pages} {3865} (\bibinfo {year}
  {1996})}\BibitemShut {NoStop}%
\bibitem [{\citenamefont {Vanderbilt}(1990)}]{uspp}%
  \BibitemOpen
  \bibfield  {author} {\bibinfo {author} {\bibfnamefont {D.}~\bibnamefont
  {Vanderbilt}},\ }\href@noop {} {\bibfield  {journal} {\bibinfo  {journal}
  {Phys. Rev. B}\ }\textbf {\bibinfo {volume} {41}},\ \bibinfo {pages} {7892}
  (\bibinfo {year} {1990})}\BibitemShut {NoStop}%
\bibitem [{\citenamefont {Montavon}\ \emph {et~al.}(2013)\citenamefont
  {Montavon}, \citenamefont {Rupp}, \citenamefont {Gobre}, \citenamefont
  {Vazquez-Mayagoitia}, \citenamefont {Hansen}, \citenamefont {Tkatchenko},
  \citenamefont {Müller},\ and\ \citenamefont {von Lilienfeld}}]{Rupp:NJP}%
  \BibitemOpen
  \bibfield  {author} {\bibinfo {author} {\bibfnamefont {G.}~\bibnamefont
  {Montavon}}, \bibinfo {author} {\bibfnamefont {M.}~\bibnamefont {Rupp}},
  \bibinfo {author} {\bibfnamefont {V.}~\bibnamefont {Gobre}}, \bibinfo
  {author} {\bibfnamefont {A.}~\bibnamefont {Vazquez-Mayagoitia}}, \bibinfo
  {author} {\bibfnamefont {K.}~\bibnamefont {Hansen}}, \bibinfo {author}
  {\bibfnamefont {A.}~\bibnamefont {Tkatchenko}}, \bibinfo {author}
  {\bibfnamefont {K.-R.}\ \bibnamefont {Müller}}, \ and\ \bibinfo {author}
  {\bibfnamefont {O.~A.}\ \bibnamefont {von Lilienfeld}},\ }\href@noop {}
  {\bibfield  {journal} {\bibinfo  {journal} {New J. Phys.}\ }\textbf {\bibinfo
  {volume} {15}},\ \bibinfo {pages} {095003} (\bibinfo {year}
  {2013})}\BibitemShut {NoStop}%
\bibitem [{\citenamefont {Friedman}\ \emph {et~al.}(2001)\citenamefont
  {Friedman}, \citenamefont {Hastie},\ and\ \citenamefont {Tibshirani}}]{elsl}%
  \BibitemOpen
  \bibfield  {author} {\bibinfo {author} {\bibfnamefont {J.}~\bibnamefont
  {Friedman}}, \bibinfo {author} {\bibfnamefont {T.}~\bibnamefont {Hastie}}, \
  and\ \bibinfo {author} {\bibfnamefont {R.}~\bibnamefont {Tibshirani}},\
  }\href@noop {} {\emph {\bibinfo {title} {The elements of statistical
  learning}}}\ (\bibinfo  {publisher} {Springer series in statistics Springer,
  Berlin},\ \bibinfo {year} {2001})\BibitemShut {NoStop}%
\bibitem [{\citenamefont {Chen}\ and\ \citenamefont
  {Guestrin}(2016)}]{xgboost}%
  \BibitemOpen
  \bibfield  {author} {\bibinfo {author} {\bibfnamefont {T.}~\bibnamefont
  {Chen}}\ and\ \bibinfo {author} {\bibfnamefont {C.}~\bibnamefont
  {Guestrin}},\ }\href@noop {} {\bibfield  {journal} {\bibinfo  {journal}
  {arXiv preprint arXiv:1603.02754}\ } (\bibinfo {year} {2016})}\BibitemShut
  {NoStop}%
\bibitem [{\citenamefont {Fink}\ \emph {et~al.}(2005)\citenamefont {Fink},
  \citenamefont {Bruggesser},\ and\ \citenamefont {Reymond}}]{fink:gdb13-1}%
  \BibitemOpen
  \bibfield  {author} {\bibinfo {author} {\bibfnamefont {T.}~\bibnamefont
  {Fink}}, \bibinfo {author} {\bibfnamefont {H.}~\bibnamefont {Bruggesser}}, \
  and\ \bibinfo {author} {\bibfnamefont {J.-L.}\ \bibnamefont {Reymond}},\
  }\href@noop {} {\bibfield  {journal} {\bibinfo  {journal} {Angew. Chem. Int.
  Ed.}\ }\textbf {\bibinfo {volume} {44}},\ \bibinfo {pages} {1504} (\bibinfo
  {year} {2005})}\BibitemShut {NoStop}%
\bibitem [{\citenamefont {Fink}\ and\ \citenamefont
  {Reymond}(2007)}]{fink:gdb13-2}%
  \BibitemOpen
  \bibfield  {author} {\bibinfo {author} {\bibfnamefont {T.}~\bibnamefont
  {Fink}}\ and\ \bibinfo {author} {\bibfnamefont {J.-L.}\ \bibnamefont
  {Reymond}},\ }\href@noop {} {\bibfield  {journal} {\bibinfo  {journal} {J.
  Chem. Inf. Model.}\ }\textbf {\bibinfo {volume} {47}},\ \bibinfo {pages}
  {342} (\bibinfo {year} {2007})}\BibitemShut {NoStop}%
\bibitem [{\citenamefont {Friedman}\ and\ \citenamefont
  {Popescu}(2008)}]{Friedman:Ensemble}%
  \BibitemOpen
  \bibfield  {author} {\bibinfo {author} {\bibfnamefont {J.~H.}\ \bibnamefont
  {Friedman}}\ and\ \bibinfo {author} {\bibfnamefont {B.~E.}\ \bibnamefont
  {Popescu}},\ }\href@noop {} {\bibfield  {journal} {\bibinfo  {journal} {Ann.
  Appl. Stat.}\ }\textbf {\bibinfo {volume} {2}},\ \bibinfo {pages} {916}
  (\bibinfo {year} {2008})}\BibitemShut {NoStop}%
\bibitem [{\citenamefont {Dai}\ \emph {et~al.}(2016)\citenamefont {Dai},
  \citenamefont {Dai},\ and\ \citenamefont {Song}}]{Dai:latent}%
  \BibitemOpen
  \bibfield  {author} {\bibinfo {author} {\bibfnamefont {H.}~\bibnamefont
  {Dai}}, \bibinfo {author} {\bibfnamefont {B.}~\bibnamefont {Dai}}, \ and\
  \bibinfo {author} {\bibfnamefont {L.}~\bibnamefont {Song}},\ }\href@noop {}
  {\bibfield  {journal} {\bibinfo  {journal} {arXiv preprint arXiv:1603.05629}\
  } (\bibinfo {year} {2016})}\BibitemShut {NoStop}%
\bibitem [{\citenamefont {Rupp}(2015)}]{Rupp:Rev}%
  \BibitemOpen
  \bibfield  {author} {\bibinfo {author} {\bibfnamefont {M.}~\bibnamefont
  {Rupp}},\ }\href@noop {} {\bibfield  {journal} {\bibinfo  {journal} {Int. J.
  Quantum Chem.}\ }\textbf {\bibinfo {volume} {115}},\ \bibinfo {pages} {1058}
  (\bibinfo {year} {2015})}\BibitemShut {NoStop}%
\bibitem [{\citenamefont {Sch\"utt}\ \emph {et~al.}(2014)\citenamefont
  {Sch\"utt}, \citenamefont {Glawe}, \citenamefont {Brockherde}, \citenamefont
  {Sanna}, \citenamefont {M\"uller},\ and\ \citenamefont
  {Gross}}]{Gross:Solids}%
  \BibitemOpen
  \bibfield  {author} {\bibinfo {author} {\bibfnamefont {K.~T.}\ \bibnamefont
  {Sch\"utt}}, \bibinfo {author} {\bibfnamefont {H.}~\bibnamefont {Glawe}},
  \bibinfo {author} {\bibfnamefont {F.}~\bibnamefont {Brockherde}}, \bibinfo
  {author} {\bibfnamefont {A.}~\bibnamefont {Sanna}}, \bibinfo {author}
  {\bibfnamefont {K.~R.}\ \bibnamefont {M\"uller}}, \ and\ \bibinfo {author}
  {\bibfnamefont {E.~K.~U.}\ \bibnamefont {Gross}},\ }\href@noop {} {\bibfield
  {journal} {\bibinfo  {journal} {Phys. Rev. B}\ }\textbf {\bibinfo {volume}
  {89}},\ \bibinfo {pages} {205118} (\bibinfo {year} {2014})}\BibitemShut
  {NoStop}%
\bibitem [{\citenamefont {Faber}\ \emph {et~al.}(2015)\citenamefont {Faber},
  \citenamefont {Lindmaa}, \citenamefont {von Lilienfeld},\ and\ \citenamefont
  {Armiento}}]{Faber:Solids}%
  \BibitemOpen
  \bibfield  {author} {\bibinfo {author} {\bibfnamefont {F.}~\bibnamefont
  {Faber}}, \bibinfo {author} {\bibfnamefont {A.}~\bibnamefont {Lindmaa}},
  \bibinfo {author} {\bibfnamefont {O.~A.}\ \bibnamefont {von Lilienfeld}}, \
  and\ \bibinfo {author} {\bibfnamefont {R.}~\bibnamefont {Armiento}},\
  }\href@noop {} {\bibfield  {journal} {\bibinfo  {journal} {Int. J. Quantum
  Chem.}\ }\textbf {\bibinfo {volume} {115}},\ \bibinfo {pages} {1094}
  (\bibinfo {year} {2015})}\BibitemShut {NoStop}%
\bibitem [{\citenamefont {Snyder}\ \emph {et~al.}(2012)\citenamefont {Snyder},
  \citenamefont {Rupp}, \citenamefont {Hansen}, \citenamefont {M\"uller},\ and\
  \citenamefont {Burke}}]{Burke:ML}%
  \BibitemOpen
  \bibfield  {author} {\bibinfo {author} {\bibfnamefont {J.~C.}\ \bibnamefont
  {Snyder}}, \bibinfo {author} {\bibfnamefont {M.}~\bibnamefont {Rupp}},
  \bibinfo {author} {\bibfnamefont {K.}~\bibnamefont {Hansen}}, \bibinfo
  {author} {\bibfnamefont {K.-R.}\ \bibnamefont {M\"uller}}, \ and\ \bibinfo
  {author} {\bibfnamefont {K.}~\bibnamefont {Burke}},\ }\href@noop {}
  {\bibfield  {journal} {\bibinfo  {journal} {Phys. Rev. Lett.}\ }\textbf
  {\bibinfo {volume} {108}},\ \bibinfo {pages} {253002} (\bibinfo {year}
  {2012})}\BibitemShut {NoStop}%
\bibitem [{\citenamefont {Lorenz}\ \emph {et~al.}(2006)\citenamefont {Lorenz},
  \citenamefont {Scheffler},\ and\ \citenamefont {Gross}}]{PES-1}%
  \BibitemOpen
  \bibfield  {author} {\bibinfo {author} {\bibfnamefont {S.}~\bibnamefont
  {Lorenz}}, \bibinfo {author} {\bibfnamefont {M.}~\bibnamefont {Scheffler}}, \
  and\ \bibinfo {author} {\bibfnamefont {A.}~\bibnamefont {Gross}},\
  }\href@noop {} {\bibfield  {journal} {\bibinfo  {journal} {Phys. Rev. B}\
  }\textbf {\bibinfo {volume} {73}},\ \bibinfo {pages} {115431} (\bibinfo
  {year} {2006})}\BibitemShut {NoStop}%
\bibitem [{\citenamefont {Lorenz}\ \emph {et~al.}(2004)\citenamefont {Lorenz},
  \citenamefont {Gross},\ and\ \citenamefont {Scheffler}}]{PES-2}%
  \BibitemOpen
  \bibfield  {author} {\bibinfo {author} {\bibfnamefont {S.}~\bibnamefont
  {Lorenz}}, \bibinfo {author} {\bibfnamefont {A.}~\bibnamefont {Gross}}, \
  and\ \bibinfo {author} {\bibfnamefont {M.}~\bibnamefont {Scheffler}},\
  }\href@noop {} {\bibfield  {journal} {\bibinfo  {journal} {Chem. Phys.
  Lett.}\ }\textbf {\bibinfo {volume} {395}},\ \bibinfo {pages} {210 }
  (\bibinfo {year} {2004})}\BibitemShut {NoStop}%
\bibitem [{\citenamefont {Khorshidi}\ and\ \citenamefont
  {A.}(2016)}]{Khorshidi}%
  \BibitemOpen
  \bibfield  {author} {\bibinfo {author} {\bibfnamefont {A.}~\bibnamefont
  {Khorshidi}}\ and\ \bibinfo {author} {\bibfnamefont {P.~A.}\ \bibnamefont
  {A.}},\ }\href@noop {} {\bibfield  {journal} {\bibinfo  {journal} {Computer
  Physics Communications}\ }\textbf {\bibinfo {volume} {207}},\ \bibinfo
  {pages} {310 } (\bibinfo {year} {2016})}\BibitemShut {NoStop}%
\bibitem [{\citenamefont {Lejaeghere}\ \emph {et~al.}(2016)\citenamefont
  {Lejaeghere}, \citenamefont {Bihlmayer}, \citenamefont {Bj{\"o}rkman},
  \citenamefont {Blaha}, \citenamefont {Bl{\"u}gel}, \citenamefont {Blum},
  \citenamefont {Caliste}, \citenamefont {Castelli}, \citenamefont {Clark},
  \citenamefont {Dal~Corso} \emph {et~al.}}]{reproducibility}%
  \BibitemOpen
  \bibfield  {author} {\bibinfo {author} {\bibfnamefont {K.}~\bibnamefont
  {Lejaeghere}}, \bibinfo {author} {\bibfnamefont {G.}~\bibnamefont
  {Bihlmayer}}, \bibinfo {author} {\bibfnamefont {T.}~\bibnamefont
  {Bj{\"o}rkman}}, \bibinfo {author} {\bibfnamefont {P.}~\bibnamefont {Blaha}},
  \bibinfo {author} {\bibfnamefont {S.}~\bibnamefont {Bl{\"u}gel}}, \bibinfo
  {author} {\bibfnamefont {V.}~\bibnamefont {Blum}}, \bibinfo {author}
  {\bibfnamefont {D.}~\bibnamefont {Caliste}}, \bibinfo {author} {\bibfnamefont
  {I.~E.}\ \bibnamefont {Castelli}}, \bibinfo {author} {\bibfnamefont {S.~J.}\
  \bibnamefont {Clark}}, \bibinfo {author} {\bibfnamefont {A.}~\bibnamefont
  {Dal~Corso}},  \emph {et~al.},\ }\href@noop {} {\bibfield  {journal}
  {\bibinfo  {journal} {Science}\ }\textbf {\bibinfo {volume} {351}},\ \bibinfo
  {pages} {aad3000} (\bibinfo {year} {2016})}\BibitemShut {NoStop}%
\end{thebibliography}%

\end{document}